%% file: B-L_couplings.tex
\documentclass[aps,prd,12pt,preprint,superscriptaddress,nofootinbib,floatfix]{revtex4}
\usepackage{latexsym}
\usepackage{amsfonts}
\usepackage[latin1]{inputenc}
\usepackage{subfig}
\usepackage{graphicx}
\usepackage{mathrsfs}
\usepackage{amssymb}
\usepackage{amsmath}
\usepackage{verbatim}
\usepackage{multirow}
\def\D0{\slash\!\!\!\!\!\!\!\!\!\:D0}

\begin{document}

\begin{flushleft}
{SHEP-11-15}\\
\today
\end{flushleft}

\title{Theoretical constraints on the couplings \\ [2mm] of non-exotic
  minimal $Z'$ bosons}
\vspace*{1.0truecm}
\author{Lorenzo Basso}
\affiliation{
School of Physics \& Astronomy, University of Southampton,\\
Highfield, Southampton SO17 1BJ, UK}
\affiliation{
Particle Physics Department, Rutherford Appleton Laboratory, \\Chilton,
Didcot, Oxon OX11 0QX, UK}
\author{Stefano Moretti}
\affiliation{
School of Physics \& Astronomy, University of Southampton,\\
Highfield, Southampton SO17 1BJ, UK}
\affiliation{
Particle Physics Department, Rutherford Appleton Laboratory, \\Chilton,
Didcot, Oxon OX11 0QX, UK}
\author{Giovanni Marco Pruna}
\affiliation{
School of Physics \& Astronomy, University of Southampton,\\
Highfield, Southampton SO17 1BJ, UK}
\affiliation{
Particle Physics Department, Rutherford Appleton Laboratory, \\Chilton,
Didcot, Oxon OX11 0QX, UK}
\begin{abstract}
{\small \noindent
We have combined perturbative unitarity and renormalisation group
equation arguments in order to find a dynamical way to constrain the
space of the gauge couplings ($g'_1$, $\widetilde{g}$) of the
so-called ``Minimal $Z'$ Models''. We have analysed the role of the 
gauge couplings evolution in the perturbative stability
of the two-to-two body scattering amplitudes of the vector and
scalar sectors of these models and we have shown that perturbative
unitarity imposes an upper bound that is generally stronger than the
triviality constraint. We have also demonstrated how this method quantitatively
refines the usual triviality bound in the case of benchmark scenarios such
as the $U(1)_\chi$, the $U(1)_R$ or the ``pure'' $U(1)_{B-L}$ extension of the
Standard Model. Finally, a description of the underlying model
structure in Feynman gauge 
is provided.}
\end{abstract}
\maketitle

\newpage


\section{Introduction}
\label{Sec:Intro}
\noindent
\input{sect_1.tex}


\section{The parametrisation of Minimal $Z'$ Models}
\label{Sec:Model}
\noindent
\input{sect_2.tex}


\section{Constraining the $g'_1-\tilde{g}$ space}
\label{Sec:bounds}
\noindent
\input{sect_3.tex}


\section{Results}
\label{Sec:Results}
\noindent
\input{sect_4.tex}


\section{Conclusions}
\label{Sec:Conclusions}
\noindent
\input{sect_5.tex}


\section*{Acknowledgements} 
\noindent
\input{acknowledgements.tex}

\newpage

\appendix
\section{Gauge-fixing Lagrangian of Minimal $Z'$ Models} 
\noindent
\input{appe_b.tex}
\label{appe:b}

\newpage
\section{Feynman rules associated with a neutral gauge boson exchange
  in a scalar two-body scattering} 
\noindent
\input{appe_a.tex}
\label{appe:a}

\newpage
\section{Explicit value of the $f_{i,j}^{z,z'}$ coefficients of
  equations~(\ref{w's})-(\ref{structure})}
\noindent
\input{appe_c.tex}
\label{appe:c}


\newpage
\bibliographystyle{h-physrev5}
\bibliography{biblio}

\end{document}

%% file: sect_1.tex
Nowadays the phenomenological importance of Beyond the Standard Model
($BSM$) physics at the TeV scale is recognised by the global
experimental effort at the Large Hadron Collider (LHC).

It is common belief that a $Z'$ boson is among the first new objects
that can potentially be detected at the LHC. The existing extensive
literature is testimonial to the growing interest in them (see
e.g.~\cite{Rizzo:1996ce,Cvetic:1997wu,Leike:1998wr,Carena:2004xs}). A
particularly interesting class of theoretical scenarios incorporating
a $Z'$ boson are the so-called ``(non-exotic/non-anomalous) 
Minimal $Z'$ Models'', extensively studied in the recent
years \cite{Appelquist:2002mw,Chankowski:2006jk,Ferroglia:2006mj,Langacker:2008yv,Erler:2009jh,Salvioni:2009mt}.

These models are based on an extension of the Standard Model ($SM$)
gauge group with a further $U(1)$ symmetry factor. The anomaly
cancellation conditions imply the inclusion of three generations of
right-handed neutrinos in the fermion sector, while the breaking of
the new gauge group is provided by an extra singlet Higgs boson
(thereby making the $Z'$ boson a massive state).

The purpose of this paper is to show that renormalisation group
equation ($RGE$) based
techniques \cite{Linde:1975sw,Linde:1975gx,Weinberg:1976pe,Wilson:1971dh,Wilson:1973jj} 
as well as a standard perturbative unitarity criterion \cite{Lee:1977eg} can
be combined to give a dynamical way to constrain the two gauge
couplings ($g'_1$ and $\widetilde{g}$) of a set of Minimal $Z'$
Models, with a particular attention devoted to some
benchmark scenarios such as the ``minimal'' $U(1)_{B-L}$, the $U(1)_R$
(no fermion charge 
associated to the left-handed fermions) and the $SO(10)$-inspired
$U(1)_{\chi}$ 
extensions (see~\cite{Carena:2004xs} for an extensive overview).

To this end, we propose a detailed study of the Goldstone and Higgs
sectors of this kind of models with a view to extract the most
stringent bounds on the (evolving) gauge couplings. 
We will make a comparison between this method and triviality
arguments, showing that calling for perturbative unitarity
stability conditions gives stronger constraints on $g'_1$ and
$\widetilde{g}$ with respect to traditional triviality assumptions
over most of the parameter space. For an exhaustive description of the
theoretical setup and of our conventions see~\cite{Basso:2010jm},
where also 
the $RGE$ equations can be found. Finally, regarding perturbative
unitarity techniques, we will expand below upon the methodology outlined
in~\cite{Basso:2010jt}. 

This work is organised as follows: in Section~\ref{Sec:Model} we
introduce our parametrisation of the Scalar Lagrangian of the Minimal
$Z'$ Models, in Section~\ref{Sec:bounds} we describe the theoretical
methods 
adopted to constrain the gauge couplings, in Section~\ref{Sec:Results}
we present our numerical results while in the last section we give our
conclusions; in Appendix~\ref{appe:b} we discuss the gauge-fixing
Lagrangian of Minimal $Z'$ Models, in Appendix~\ref{appe:a} we list
the set of Feynman rules that is relevant in our calculation, and in
Appendix~\ref{appe:c} we give some explicit analytical results that
have been used in this paper.

%% file: sect_2.tex
We describe here our parametrisation of the Minimal $Z'$
Models. Following~\cite{Basso:2010jm}, the $SM$ gauge group is
augmented by a $U(1)$ factor, 
related to the Baryon minus Lepton ($B-L$) gauged number. In the
complete model, the classical gauge invariant Lagrangian, obeying the
$SU(3)_C\times SU(2)_L\times U(1)_Y\times U(1)_{B-L}$ gauge symmetry,
can be decomposed as:
\begin{equation}\label{L}
\mathscr{L}=\mathscr{L}_s + \mathscr{L}_{YM} + \mathscr{L}_f + \mathscr{L}_Y \, .
\end{equation}

In this paper we are mainly interested in the scalar part of the
Lagrangian 
\begin{equation}\label{new-scalar_L}
\mathscr{L}_s=\left( D^{\mu} H\right) ^{\dagger} D_{\mu}H + 
\left( D^{\mu} \chi\right) ^{\dagger} D_{\mu}\chi - V(H,\chi ) \, ,
\end{equation}
with the scalar potential given by
\begin{eqnarray}\nonumber
V(H,\chi )&=&m^2H^{\dagger}H + \mu ^2\mid\chi\mid ^2
+\left( \begin{array}{cc} H^{\dagger}H& \mid\chi\mid
  ^2\end{array}\right) \left( \begin{array}{cc} \lambda _1 &
    \frac{\lambda _3}{2} \\ \frac{\lambda _3}{2} & \lambda _2
    \\ \end{array} \right) \left( \begin{array}{c} H^{\dagger}H
    \\ \mid\chi\mid ^2 \\ \end{array} \right)\\ \nonumber
  \\ \label{BL-potential} &=&m^2H^{\dagger}H + \mu ^2\mid\chi\mid ^2 +
  \lambda _1 (H^{\dagger}H)^2 +\lambda _2 \mid\chi\mid ^4 + \lambda _3
  H^{\dagger}H\mid\chi\mid ^2  \, , 
\end{eqnarray}
where $H$ and $\chi$ are the complex scalar Higgs doublet and singlet
fields, respectively.

We generalise the $SM$ discussion of spontaneous Electro-Weak Symmetry
Breaking ($EWSB$) to the more
complicated classical potential of equation~(\ref{BL-potential}). To
determine the conditions for $V(H,\chi )$ to be bounded from below, it
is sufficient to study its behaviour for large field values,
controlled by the matrix in the first line of
equation~(\ref{BL-potential}). Requiring such a matrix to be
positive-definite, we obtain the conditions:
\begin{equation}\label{inf_limitated}
4 \lambda _1 \lambda _2 - \lambda _3^2>0 \, ,
\end{equation}
\begin{equation}\label{positivity}
\lambda _1, \lambda _2 > 0 \, .
\end{equation}

If the above conditions are satisfied, we can proceed to the
minimisation of $V$ as a function of constant Vacuum Expectation
Values ($VEVs$) for the two Higgs fields. 
In the Feynman gauge, we can parametrise the scalar fields as
\begin{equation}\label{Higgs_goldstones}
H=\frac{1}{\sqrt{2}}
\left(
\begin{array}{c}
-i(\phi_1-i\phi_2) \\
v+(h+i\phi_3)
\end{array}
\right), \qquad
\chi =
\frac{1}{\sqrt{2}}
\left(x+(h'+i\phi_4)\right),
\end{equation}
where $w^{\pm}=\phi_1 \mp i\phi_2$ are the would-be Goldstone bosons
of $W^{\pm}$, while $\phi_3$ and $\phi_4$ will mix to give $z$ and
$z'$, the would-be Goldstone bosons of the $Z$ and $Z'$ bosons,
respectively. The real and non-negative $VEVs$ are $v$ and $x$, for
the Higgs doublet and singlet, respectively.

%

We denote by $h_1$ and $h_2$ the scalar fields of definite masses,
$m_{h_1}$ and $m_{h_2}$ respectively, and we conventionally choose
$m^2_{h_1} < m^2_{h_2}$. After standard manipulations, the explicit
expressions for the scalar mass eigenvalues and eigenvectors are:
\begin{eqnarray}\label{mh1}
m^2_{h_1} &=& \lambda _1 v^2 + \lambda _2 x^2 - \sqrt{(\lambda _1 v^2
  - \lambda _2 x^2)^2 + (\lambda _3 xv)^2} \, ,\\ \label{mh2} 
m^2_{h_2} &=& \lambda _1 v^2 + \lambda _2 x^2 + \sqrt{(\lambda _1 v^2
  - \lambda _2 x^2)^2 + (\lambda _3 xv)^2} \, , 
\end{eqnarray}
\begin{equation}\label{scalari_autostati_massa}
\left( \begin{array}{c} h_1\\
h_2\end{array}\right) =
  \left( \begin{array}{cc}
    \cos{\alpha}&-\sin{\alpha}
\\ \sin{\alpha}&\cos{\alpha} 
	\end{array}\right) \left( \begin{array}{c}h
\\h'\end{array}\right) \, , 
\end{equation}
where $-\frac{\pi}{2}\leq \alpha \leq \frac{\pi}{2}$
fulfils\footnote{In all generality, the whole interval $0\leq \alpha <
  2\pi$ is halved because an orthogonal transformation is invariant
  under $\alpha \rightarrow \alpha + \pi$. We could re-halve the
  interval by noting that it is invariant also under $\alpha
  \rightarrow -\alpha$ if we permit the eigenvalues inversion, but
  this is forbidden by our convention $m^2_{h_1} < m^2_{h_2}$. Thus
  $\alpha$ and $-\alpha$ are independent
  solutions.}:\label{scalar_angle} 
\begin{eqnarray}\label{sin2a}
\sin{2\alpha} &=& \frac{\lambda _3 xv}{\sqrt{(\lambda _1 v^2 - \lambda
    _2 x^2)^2 + (\lambda _3 xv)^2}} \, ,\\ \label{cos2a} 
\cos{2\alpha} &=& \frac{\lambda _1 v^2 - \lambda _2
  x^2}{\sqrt{(\lambda _1 v^2 - \lambda _2 x^2)^2 + (\lambda _3
    xv)^2}}\, . 
\end{eqnarray}

For our numerical study of the extended Higgs sector, it is useful to
invert equations~(\ref{mh1}), (\ref{mh2}) and (\ref{sin2a}) to extract the
parameters in the Lagrangian in terms of the physical quantities
$m_{h_1}$, $m_{h_2}$ and $\alpha$: 
\begin{eqnarray}\nonumber
\lambda _1 &=& \frac{m_{h_2}^2}{4v^2}(1-\cos{2\alpha}) +
\frac{m_{h_1}^2}{4v^2}(1+\cos{2\alpha}),\\ \nonumber 
\lambda _2 &=& \frac{m_{h_1}^2}{4x^2}(1-\cos{2\alpha}) +
\frac{m_{h_2}^2}{4x^2}(1+\cos{2\alpha}),\\ \label{inversion} 
\lambda _3 &=& \sin{2\alpha} \left( \frac{m_{h_2}^2-m_{h_1}^2}{2xv}
\right). 
\end{eqnarray}

In order to determine the covariant derivative, we must introduce 
$\mathscr{L}_{YM}$, in which the the non-Abelian field strengths 
therein are the same as in the $SM$ whereas the Abelian ones can be
written as follows: 
\begin{equation}\label{La}
\mathscr{L}^{\rm Abel}_{YM} = 
-\frac{1}{4}F^{\mu\nu}F_{\mu\nu}-\frac{1}{4}F^{\prime\mu\nu}F^\prime
_{\mu\nu}\, , 
\end{equation}
where
\begin{eqnarray}\label{new-fs3}
F_{\mu\nu} &=& \partial _{\mu}B_{\nu} - \partial _{\nu}B_{\mu} \, ,
\\ \label{new-fs4} 
F^\prime_{\mu\nu}	&=& \partial _{\mu}B^\prime_{\nu} - \partial
_{\nu}B^\prime_{\mu} \, . 
\end{eqnarray}

In this field basis, the covariant derivative is:
\begin{equation}\label{cov_der}
D_{\mu}\equiv \partial _{\mu} + ig_S
T^{\alpha}G_{\mu}^{\phantom{o}\alpha}  + igT^aW_{\mu}^{\phantom{o}a}
+ig_1YB_{\mu} +i(\widetilde{g}Y + g_1'Y_{B-L})B'_{\mu}\, . 
\end{equation}

To determine the boson spectrum, we have to expand the scalar
kinetic terms like for the $SM$. As for the gauge bosons, we expect that
there exists a mass-less gauge boson, the photon, whilst the other
gauge bosons become massive. The extension we are studying is in the
Abelian sector of the $SM$ gauge group, so that the charged gauge
bosons $W^\pm$ will have masses given by their usual $SM$ expressions,
being related to the $SU(2)_L$ factor only. The gauge boson spectrum
is then extracted from the kinetic terms in
equation~(\ref{new-scalar_L}):
\begin{eqnarray}\nonumber
\left. \left( D^{\mu} H\right) ^\dagger
D_{\mu}H \right|_{gauge}&=& \frac{1}{2}\partial 
^{\mu} h \partial _{\mu}h + \frac{1}{8} (h+v)^2 \big( 0\; 1 \big)
\Big[ g W_a ^{\phantom{o}\mu }\sigma _a + g_1B^{\mu}+\widetilde g
  B'^{\mu} \Big] ^2 \left( \begin{array}{c} 0\\1\end{array} \right)
  \\ \nonumber &=& \frac{1}{2}\partial ^{\mu} h \partial _{\mu}h +
  \frac{1}{8} (h+v)^2 \left[ g^2 \left| W_1 ^{\phantom{o}\mu } - iW_2
    ^{\phantom{o}\mu } \right| ^2  \right.\\ \label{boson_masses1} &&
    \left. \hspace{4cm} + \left( gW_3 ^{\phantom{o}\mu } - g_1 B^{\mu}
    - \widetilde g B'^{\mu}\right) ^2 \right] \, , 
\end{eqnarray}
and
\begin{eqnarray}\label{boson_masses2}
\left. \left( D^{\mu} \chi\right) ^\dagger D_{\mu}\chi \right|_{gauge}
&=& \frac{1}{2}\partial ^{\mu} h' \partial _{\mu}h'
+ \frac{1}{2}(h'+x)^2 (g_1' 2B'^{\mu})^2\, ,
\end{eqnarray}
where we have taken $Y^{B-L}_\chi = 2$ in order to guarantee the gauge
invariance of the Yukawa terms
(see \cite{Jenkins:1987ue,Buchmuller:1991ce} for details). In
equation~(\ref{boson_masses1}) we can recognise the $SM$ charged
gauge bosons $W^\pm$, with $\displaystyle M_W=gv/2$ as in the
$SM$. The other gauge boson masses are not so simple to identify,
because of mixing. In fact, in analogy with the $SM$, the fields of
definite mass are linear combinations of $B^\mu$, $W_3^\mu$ and
$B'^\mu$. The explicit expressions are: 
\begin{equation}\label{neutral_bosons}
\left( \begin{array}{c} B^{\mu}
  \\ W_3^{\phantom{o}\mu}\\ B'^{\mu} \end{array}\right) =
\left( \begin{array}{ccc} \cos{\vartheta _w} & -\sin{\vartheta
    _w}\cos{\vartheta '} & \sin{\vartheta _w}\sin{\vartheta
    '}\\ \sin{\vartheta _w} & \cos{\vartheta _w}\cos{\vartheta '} &
  -\cos{\vartheta _w}\sin{\vartheta '}\\ 0 & \sin{\vartheta '} &
  \cos{\vartheta '} \end{array} \right) \left( \begin{array}{c}
  A^{\mu} \\ Z^{\mu}\\ Z'^{\mu} \end{array}\right)\, ,\\ 
\end{equation}
with $-\frac{\pi}{4}\leq \vartheta '\leq \frac{\pi}{4}$, such that: 
\begin{equation}\label{tan2theta_prime}
\tan{2\vartheta
  '}=\frac{2\widetilde{g}\sqrt{g^2+g_1^2}}{\widetilde{g}^2 +
  16(\frac{x}{v})^2 g_1^{'2}-g^2-g_1^2} 
\end{equation}
and
\begin{eqnarray}\nonumber
M_A &=& 0\, ,\\ \label{Mzz'}
M_{Z,Z'}^2 &=& \frac{1}{8}\left(C v^2 \mp \sqrt{-D+v^4 C^2}\right)\, , 
\end{eqnarray}

where
\begin{eqnarray}
	  C &=& g^2 + g_1^2+\widetilde{g}^2 +
	  16 \left( \frac{x}{v}\right)^2 g_1^{'2}\, ,\\
	  D &=& 64v^2x^2(g^2 + g_1^2)g_1^{'2} \, .
\end{eqnarray}

As for the Goldstone boson spectrum, it is possible to find a
convenient way to write the mass matrix. Being $H\sim
(1,2,\frac{1}{2},0)$ and $\chi\sim (1,1,0,2)$ the Higgs representations
associated to each gauge group, in the gauge-Goldstone\footnote{The
$5\times 4$ matrix follows from the five gauge bosons $W^i|_{i=1,3}$,
$Z$, $Z'$ and the four Goldstone bosons $\phi^i|_{i=1,4}$.} bosons
basis we find the following representation of the co-variant
derivative:
\begin{equation}\label{gF}
\mathcal{D} = \left(
\begin{array}{cccc}
\frac{v}{2}g	&0	&0	&0\\
0&	\frac{v}{2}g	&0	&0\\
0&	0&\frac{v}{2}g	&0\\
0&	0&-\frac{v}{2}g_1	&0\\
0&	0&-\frac{v}{2}\widetilde{g}	&-2xg'_1\\
\end{array} \right)\, .
\end{equation}

In the t'Hooft-Feynman gauge, it can be verified that the vector boson
mass matrix is given by $m^2_V = \mathcal{D} (\mathcal{D})^T$. The
related Goldstones mass matrix can as well be evaluated as
\begin{equation}
m^2_v = (\mathcal{D})^T \mathcal{D}\, ,
\end{equation}
therefore we get
\begin{equation}\label{GB_mass_matrix}
m^2_v = \left(
\begin{array}{cccc}
\frac{v^2}{4}g^2	&0	&0	&0\\
0&\frac{v^2}{4}g^2	&0	&0\\
0&	0&\frac{v^2}{4}(g^2+g1^2+\widetilde{g}^2) & xv\widetilde{g}g'_1\\
0&	0&xv\widetilde{g}g'_1&4x^2(g'_1)^2\\
\end{array} \right)\, .
\end{equation}

The mass matrix in equation~(\ref{GB_mass_matrix}) shows that the
Goldstones of the $W$-boson have a mass that is equivalent to the $SM$
one, while the $\phi_3$ and $\phi_4$ fields mix, as it happens for the
related gauge bosons. We can diagonalise the neutral Goldstone block
by means of a rotation of angle $\alpha_g$, defined by: 
\begin{equation}
\tan{2\alpha _g}
= \frac{-8 \frac{x}{v} \widetilde{g} \, g'_1}{g^2 + g_1^2
+ \widetilde{g}^2 - 16 \left(\frac{x}{v}g'_1\right)^2}\, ,
\end{equation}
obtaining, as expected, the neutral gauge boson masses as eigenvalues
of the neutral Goldstone boson sub-matrix. As for the neutral gauge
boson sector, the Goldstones mix only if $\widetilde{g}\neq
0$. Finally, the neutral Goldstone bosons fulfil 
\begin{equation}\label{goldstones_autostati_massa}
\left( \begin{array}{c} z\\z'\end{array}\right) =
\left( \begin{array}{cc} \cos{\alpha_g}& \sin{\alpha_g}\\
-\sin{\alpha_g}&\cos{\alpha_g} 
	\end{array}\right) \left( \begin{array}{c} \phi_3\\
\phi_4\end{array}\right) \,.
\end{equation}

Now that the scalar Lagrangian has been presented in the Feynman
gauge, we have all the elements to carry on with our
analysis. Although not relevant for the latter, we also present for
completeness the gauge-fixing Lagrangian in
Appendix~\ref{appe:b}.




The generic model that has been previously introduced spans over a
continuous set of minimal $U(1)$ extensions of the $SM$, that can be
labelled by the properties of the charge assignments to the particle
content.

As any other parameter in the Lagrangian, $\widetilde{g}$ and $g_1'$
are running parameters, therefore their values have to be set at some
scale. A discrete set of popular $Z'$ models (see, e.g.~\cite{Carena:2004xs,Appelquist:2002mw}) can be recovered by
a suitable definition of both $\widetilde{g}$ and $g_1'$. 

Even though we present results in the generic ($g_1'$--$\widetilde{g}$) space, we will comment
on a subset of particular interest: the
``pure'' $B-L$ extension $U(1)_{B-L}$
($\widetilde{g}_{EW}=0$) has a vanishing mixing between the massive
neutral gauge bosons at tree-level at the $EW$ scale, the
$SO(10)$-inspired extension $U(1)_{\chi}$
($\widetilde{g}_{EW}=-4/5g'_1$) preserves the mixing ratio at any
energy scale and the $R$ minimal extension $U(1)_{R}$
($\widetilde{g}_{EW}=-2g'_1$) in which a $Z'$ is coupled to
right-handed fermions only.
%
%
%
In Table~\ref{models} we summarise these models emerging from our parametrisation.

\begin{table}[!ht]
\begin{center}
\begin{tabular}{|c|l|}
\hline
Model & Parametrisation \\
\hline
$U(1)_{B-L}$ & $\widetilde{g}_{EW}=0$ \\
\hline
$U(1)_\chi$ & $\widetilde{g}_{EW}=-4/5g'_1$ \\ 
\hline
$U(1)_R$ & $\widetilde{g}_{EW}=-2g'_1$  \\ 
\hline
\end{tabular}
\end{center}
\caption{Specific parametrisations of the Minimal $Z'$ Models:
$U(1)_{B-L}$, $U(1)_\chi$ and $U(1)_R$.}
\label{models}
\end{table}

%% file: sect_3.tex
Since it has been proven that perturbative unitarity violation at high 
energy occurs only in 
vector and Higgs boson elastic scatterings, our 
interest is focused on the corresponding sectors that have been
already presented in Section~\ref{Sec:Model}.

Following the Becchi-Rouet-Stora ($BRS$) invariance
(see \cite{Becchi:1975nq}),  the amplitude for 
emission or absorption of a ``scalarly'' polarised gauge
boson becomes equal to the amplitude 
for emission or absorption of the related would-be-Goldstone boson,
and, in the high energy limit ($s \gg m^2_{W^{\pm},Z,Z'}$), the
amplitude involving the (physical) longitudinal polarisation (the
dominant one at high energies) of 
gauge bosons approaches the (unphysical) scalar one, proving the so-called
Equivalence Theorem ($ET$), see \cite{Chanowitz:1985hj}. Therefore, the
analysis of the 
perturbative unitarity of two-to-two
particle scatterings in the gauge sector can be performed, in the high
energy limit, by exploiting the Goldstone sector instead (further details of
this formalism can be found in \cite{Basso:2010jt}).


Moreover, since we want to focus on $g'_1$ and $\widetilde{g}$ limits,
we assume that the two Higgs bosons of the model have masses such that no 
significant contribution to the spherical partial wave amplitude (see below) will come 
from the scalar four-point and three-point functions (that is
$m_{1,2}\ll 700$ according to \cite{Basso:2010jt}), i.e. the Higgs masses
are well below the Lee-Quigg-Tacker ($LQT$) limit \cite{Lee:1977eg}. 
It is important to remark that relatively high values of the Higgs masses, far below the unitarity limit, tend to lead to quartic coupling to values that become non-perturbative at high scales. On a side, this could considerably refine the unitarity bounds. On the other side, it could be non-consistent by triviality arguments (as a general rule, the larger the cut-off, the smaller the acceptable value of the Higgs mass). Beyond any doubt, given a cut-off energy, a good choice for the Higgs masses is the one explored in \cite{Basso:2010jm}.
With
this choice we exclude any other source of unitarity violation
different from the size of the $g'_1$ and $\widetilde{g}$ gauge
couplings. 

Firstly, we focus on the techniques that
we have used to obtain the aforementioned unitarity
bounds in combination with the $RGE$ analysis: for this, it is crucial
to define the evolution of the gauge couplings via the $RGEs$ and
their boundary conditions.
As already established in \cite{BL_master_thesis,Basso:2010jm}, the
$RGEs$ of $g$, $g_1$,
$g'_1$ and $\widetilde{g}$ are:
\begin{eqnarray}\label{RGE_g}
\frac{d(g)}{d(\log{\Lambda})} &=& \frac{1}{16\pi
^2}\left[-\frac{19}{6}g^3 \right]\, , \nonumber \\
\label{RGE_g1}
\frac{d(g_1)}{d(\log{\Lambda})} &=& \frac{1}{16\pi
^2}\left[\frac{41}{6}g_1^3 \right]\, , \nonumber \\ \label{RGE_g2} 
\frac{d(g_1')}{d(\log{\Lambda})} &=& \frac{1}{16\pi
^2}\left[ 12 g_1'^3 +
2 \frac{16}{3} g_1'^2\widetilde g+\frac{41}{6}g_1'\widetilde
g^2 \right] \, ,\nonumber \\ \label{RGE_g_tilde} 
\frac{d(\widetilde g)}{d(\log{\Lambda})}&=& \frac{1}{16\pi
^2}\left[\frac{41}{6}\widetilde{g}\,(\widetilde
g^2+2g_1^2) + 2 \frac{16}{3} g_1' (\widetilde{g}^2 + g_1^2)
+ 12 g_1'^2\widetilde
g \right]\, , 
\end{eqnarray}
where $g(EW)\simeq 0.65$ and $g_1(EW)\simeq 0.36$.
This fully fixes the evolution of $g'_1$ and $\widetilde{g}$ with the
energy scale $\Lambda$.

In the search for the maximum $g'_1(EW)$ and $\widetilde{g}(EW)$ values allowed
by theoretical constraints, the contour condition
\begin{eqnarray}\label{triviality_condition}
g'_1(\Lambda),\widetilde{g}(\Lambda)\leq k,
\end{eqnarray}
also known as the triviality condition, is the assumption that enables one
to solve the above system of equations and gives the traditional upper bound on
the $g'_1$-$\widetilde{g}(EW)$ space at the $EW$ scale.

It is usually assumed either $k=1$ or $k=\sqrt{4\pi}$, calling for a
coupling that preserves the perturbative convergence of the theory.
Nevertheless, we stress again that this is an ``ad hoc''
assumption. 
Our aim, instead, is to extract the boundary conditions by perturbative
unitarity arguments, showing that, under certain conditions, it
represents a stronger constraint on most of the 
gauge couplings parameter space.
For this, we exploit the theoretical techniques that are related with
the perturbative unitarity analysis, since they can be used to provide
constraints on the theory, with a procedure that is not far from the
one firstly described in detail by \cite{Lee:1977eg}.

A well known result is that, by evaluating the tree-level scattering
amplitude of longitudinally polarised vector bosons, one finds that the
latter grows with the energy of the process and, in order to preserve
unitarity, it is necessary to include some other (model dependent)
interactions (for example, in the $SM$ one needs to add
the Higgs boson) and these must fulfil the unitarity criterion (again
in the $SM$, the Higgs boson must have a mass bounded 
from above by the $LQT$ limit \cite{Luscher:1988gk}).

As already intimated, we also know that the $ET$ allows one
to compute the amplitude of any process with external longitudinal
vector bosons $V_L$ ($V = W^\pm,Z,Z' $), in the limit $m^2_V\ll s$,
by substituting each one of these with the related Goldstone boson $v
= w^\pm,z,z'$, and its general validity has been proven
in \cite{Chanowitz:1985hj}. Schematically, if we consider a
process with four longitudinal vector bosons, we have that $M(V_L
V_L \rightarrow V_L V_L) = M(v v \rightarrow v v)+ O(m_V^2/s)$.

While in the search for the Higgs boson mass bound it is widely
accepted to assume small values for the gauge couplings and large
Higgs boson masses, for our purpose we reverse such argument with the
same logic: we assume that the Higgs boson masses are compatible with
the unitarity limits and we study the two-to-two scattering amplitudes
of the whole scalar sector, pushing the size of $g'_1$ and
$\widetilde{g}$ up to the unitarisation limit.

This limit is a consequence of the following argument: given a
tree-level scattering amplitude between two spin-$0$ particles,
$M(s,\theta)$, where $\theta$ is the scattering (polar) angle, 
we know that the partial wave amplitude with angular
momentum $J$ is given by
\begin{eqnarray}\label{integral}
a_J = \frac{1}{32\pi} \int_{-1}^{1} d(\cos{\theta}) P_J(\cos{\theta})
M(s,\theta),
\end{eqnarray}
where $P_J$ are Legendre polynomials, and it has been proven
(see \cite{Luscher:1988gk}) that, in order to preserve unitarity, each
partial wave must be bounded by the condition
\begin{eqnarray}\label{condition}
|\textrm{Re}(a_J(s))|\leq \frac{1}{2}.
\end{eqnarray}

By direct computation, it turns out that only $J=0$ (corresponding to
the spherical partial wave contribution) leads to some bound, so we
will not discuss the higher partial waves any further.

Assuming that the Higgs boson masses do not play any role in the
perturbative unitarity violation, we have verified that the only divergent 
contribution to the spherical amplitude is due to the size
of the $g'_1$ and $\widetilde{g}$ couplings in the $t$-channel
intermediate $Z$ and $Z'$ vector boson exchange contributions. 
In Appendix~\ref{appe:a} we list the relevant 3-point Feynman rules
that connect any of the two (external) scalars with either a $Z$ or
$Z'$ (mediator).
Hence, the relevant channels are represented by the 6-dimensional (symmetric)
scattering matrix in Table~\ref{scattering-matrix} plus the decoupled
eigenchannel $w^+w^-\rightarrow w^+w^-$.

\begin{table}[!ht]
\begin{center}
\begin{tabular}{|c||c|c|c|c|c|c|}
\hline
\ & $z^{\phantom{\prime}}z^{\phantom{\prime}}$ &
$z^{\phantom{\prime}}z'$ & $z'z'$ & $h_1h_1$ & $h_1h_2$ & $h_2h_2$ \\ 
\hline \hline
$z^{\phantom{\prime}}z^{\phantom{\prime}}$ & $0$ & $0$ & $0$ & $\sim$
& $\sim$ & $\sim$ \\
\hline
$z^{\phantom{\prime}}z'$ & $0$ & $0$ & $0$ & $\sim$ & $\sim$ & $\sim$ \\
\hline
$z'z'$ & $0$ & $0$ & $0$ & $\sim$ & $\sim$ & $\sim$ \\
\hline
$h_1h_1$ & $\sim$ & $\sim$ & $\sim$ & $0$ & $0$ & $0$ \\
\hline
$h_1h_2$ & $\sim$ & $\sim$ & $\sim$ & $0$ & $0$ & $0$ \\
\hline
$h_2h_2$ & $\sim$ & $\sim$ & $\sim$ & $0$ & $0$ & $0$ \\
\hline
\end{tabular}
\end{center}
\caption{Scattering matrix: we have used the simbol $\sim$ just for
illustrating the presence of a non-zero element in the correspondent
scattering channels.}
\label{scattering-matrix}
\end{table}

After explicit evaluation, the spherical amplitude of the
decoupled $w^+w^-$ eigenchannel, in the high energy limit, is:
\begin{eqnarray}\label{w's}
a_{w^+w^-}=\left\{ \frac{f^{z}_{w^+w^-}}{16\pi}\left[ 1+4\log{\left(
  \frac{M_Z}{\Lambda} \right)} \right]
+\frac{f^{z'}_{w^+w^-}}{16\pi}\left[
  1+4\log{\left( \frac{M_{Z'}}{\Lambda}\right)} \right] \right\},
\end{eqnarray}
and each element of the
scattering matrix presents the following structure:
\begin{eqnarray}\label{structure}
a_{ij}=S_iS_j\left\{ \frac{f^{z}_{i,j}}{16\pi}\left[ 1+4\log{\left(
  \frac{M_Z}{\Lambda} \right)} \right]
+\frac{f^{z'}_{i,j}}{16\pi}\left[
  1+4\log{\left( \frac{M_{Z'}}{\Lambda}\right)} \right] \right\},
\end{eqnarray}
where $S$ is a symmetry factor that becomes $1/\sqrt{2}$ if the (initial or final) state
has identical particles, $1$ otherwise, and $\Lambda$ represents the
scale of energy at which the scattering is consistent with
perturbative unitarity, i.e. it is
the evolution energy scale cut-off. It is important to notice
that the masses of the $Z$ and $Z'$ act as a natural 
regulator that preserves both the amplitude and the spherical
partial wave from any $t$-channel collinear divergence and that both
of them are
completely defined by the choice of the gauge couplings and
$VEVs$ (see equation~(\ref{Mzz'})). The non-vanishing
coefficients of equations~(\ref{w's})-(\ref{structure}) are listed in
Appendix~\ref{appe:c}.

It is well-known\footnote{The diagonalisation of the scattering matrix usually leads to stronger bounds not only in the $SM$-case but also in $BSM$ scenarios (e.g.~\cite{Kanemura:1993hm}).} that the most stringent unitarity bounds on the $g'_1$-$\widetilde{g}$ space 
are derived from the requirement that the magnitude of the largest
eigenvalue of the scattering matrix does not exceed $1/2$.

Finally, if we consider the contour of this inequality, we find
exactly the boundary conditions that solve the set of differential
equations in (\ref{RGE_g2}), giving us the upper limits for
$g'_1$ and $\widetilde{g}$ at the $EW$ scale.
In the next section we will combine all these elements to present a 
numerical analysis of the allowed domain of the gauge couplings.

%% file: sect_4.tex
The set of differential equations~(\ref{RGE_g2}) has been integrated
with the well-known Runge-Kutta algorithm and both the unitarity
(equation~(\ref{condition})) and
triviality (equation~(\ref{triviality_condition})) conditions have
been imposed as a two-point boundary
value with a simple shooting method, that consisted in varying the
initial gauge coupling values in dichotomous-converging steps until
the bounds were fulfilled.

Apart from the gauge couplings, other parameters play a role in the
computation: the $VEVs$ have been chosen in such a way that both
$M_Z$ and $M_{Z'}$ are within the allowed experimental range (see
\cite{Nakamura:2010zzi} and \cite{Cacciapaglia:2006pk}, respectively),
and further that 
$M_{Z'}$ lies in the $\mathcal{O}(1-10)$ TeV range, so that it does
not spoil the high energy approximation $M_{Z'}\ll \Lambda$. By direct
computation we verified that the Higgs mixing angle $\alpha$ does not
play any significant role in the analysis, hence for each analysed
point of the gauge couplings parameter space we have averaged
the spherical wave greatest eigenvalue over the range
$-1<\sin{\alpha}<+1$, finding a standard deviation never greater than
$\sim 2\%$. Finally, for illustrative purposes, we have chosen the
triviality condition to be fixed by $k=1$.

As initial step of our numerical analysis, we have verified by direct
computation that the spherical 
wave associated to the decoupled eigenchannel $w^+w^-\rightarrow
w^+w^-$ gives always a negligible contribution with respect to the
greatest eigenvalue of the spherical wave scattering matrix in
Table~\ref{scattering-matrix}.
\begin{figure}[!t]
\centering
  \subfloat[]{ 
  \label{cont_11}
  \includegraphics[angle=0,width=0.48\textwidth ]{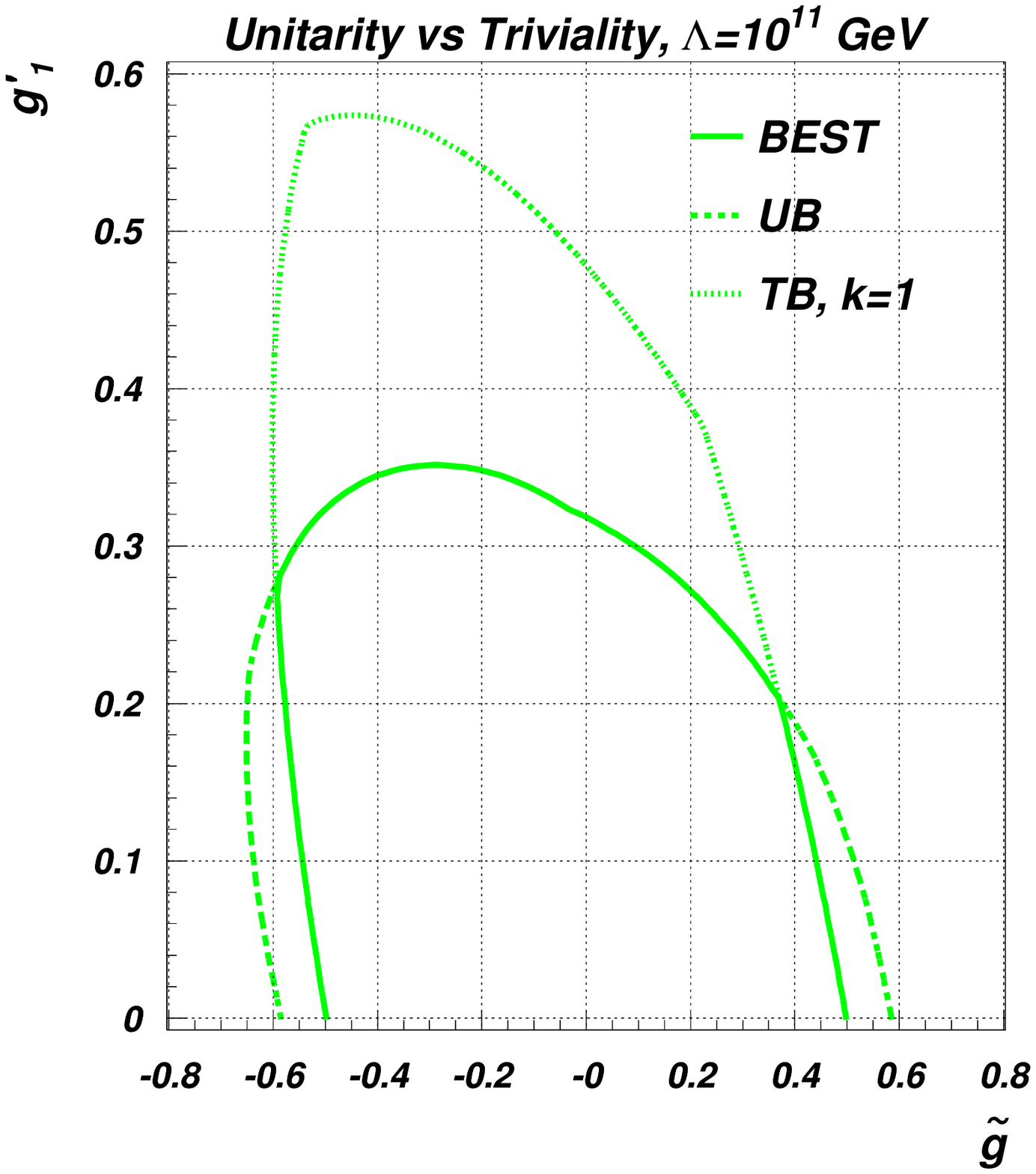}}
  \subfloat[]{ 
  \label{cont_15}
  \includegraphics[angle=0,width=0.48\textwidth ]{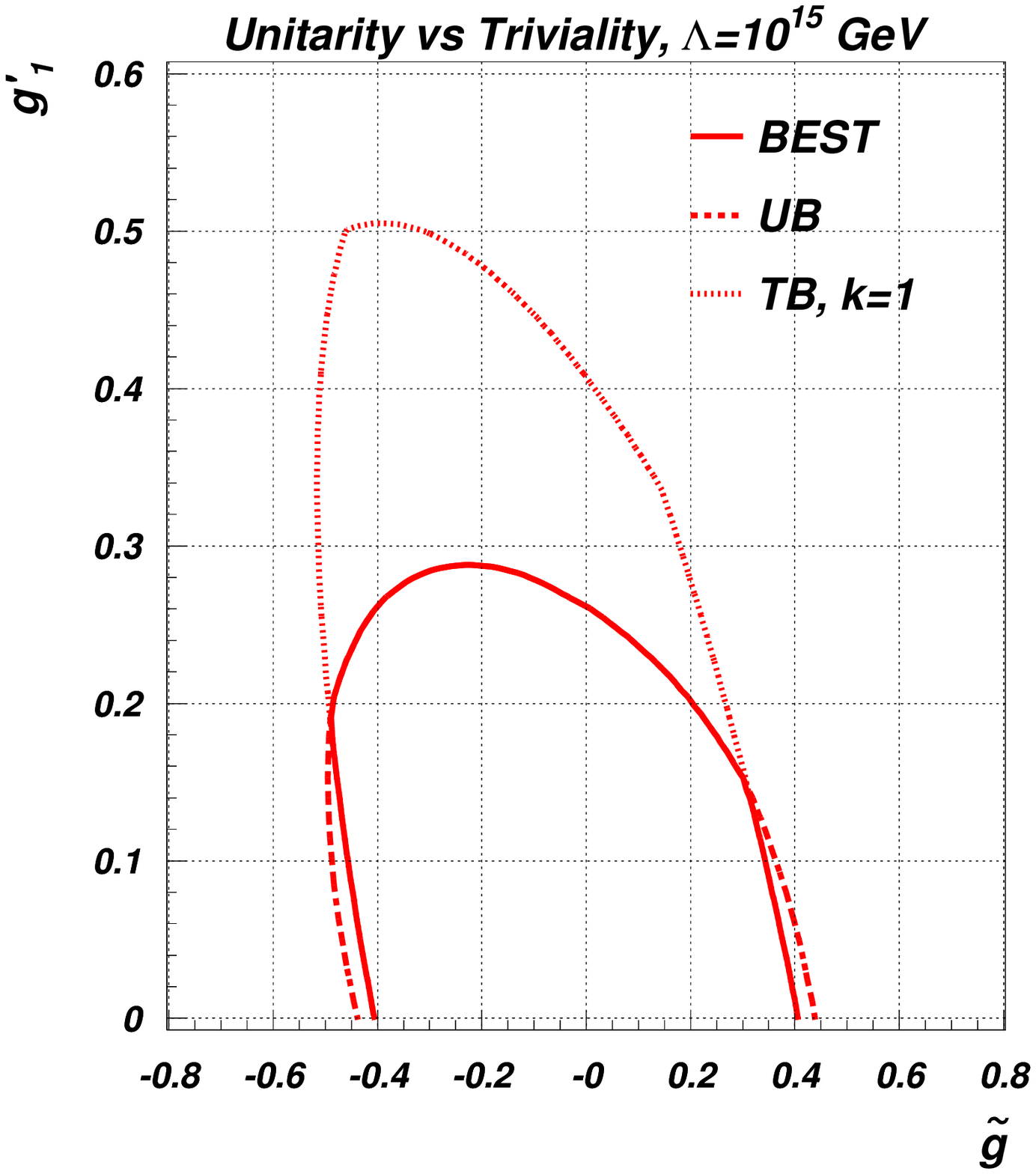}}
  \\
  \subfloat[]{
\label{cont_19}
  \includegraphics[angle=0,width=0.48\textwidth ]{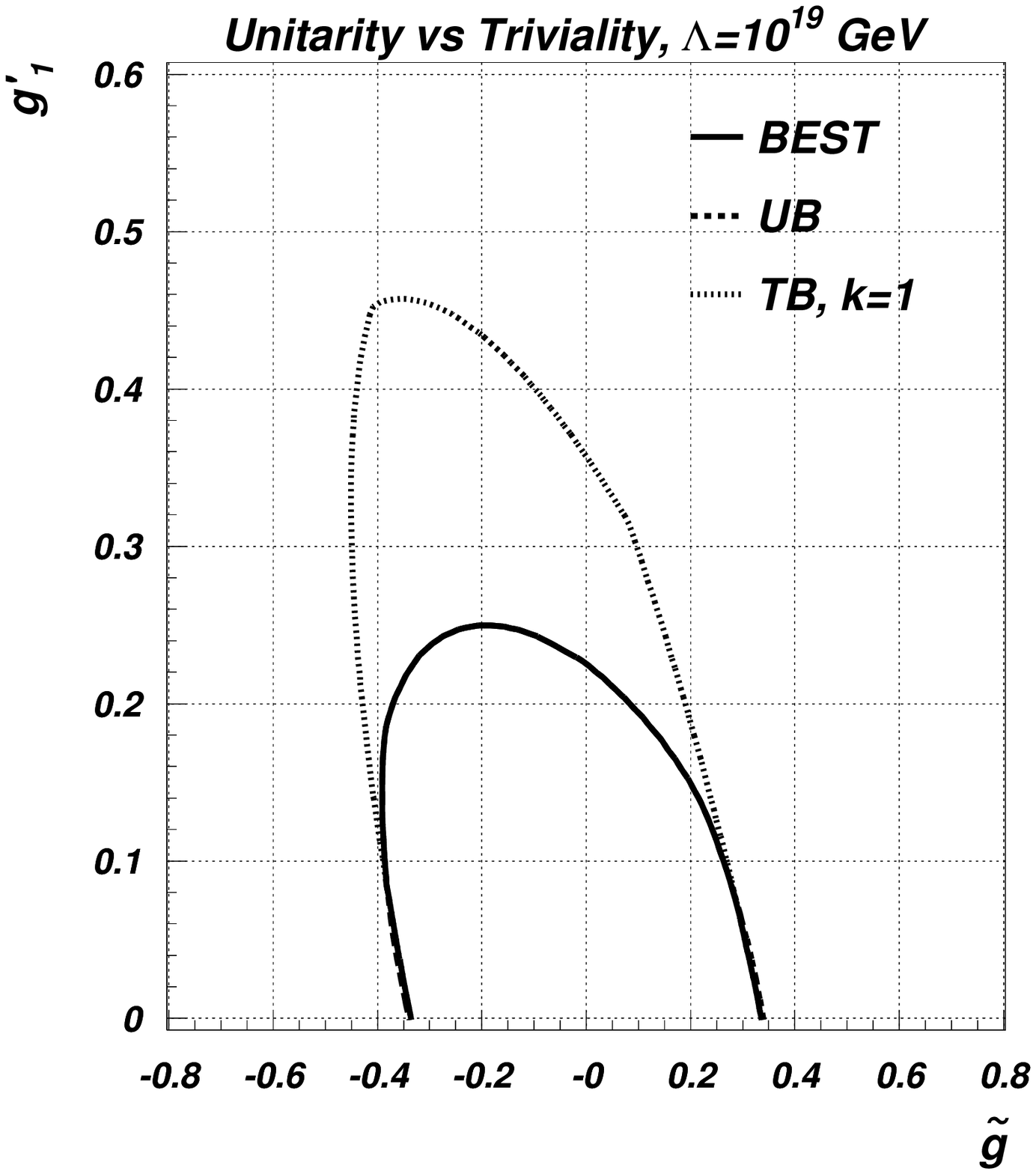}}
  \subfloat[]{
  \label{cont_all}
  \includegraphics[angle=0,width=0.48\textwidth
  ]{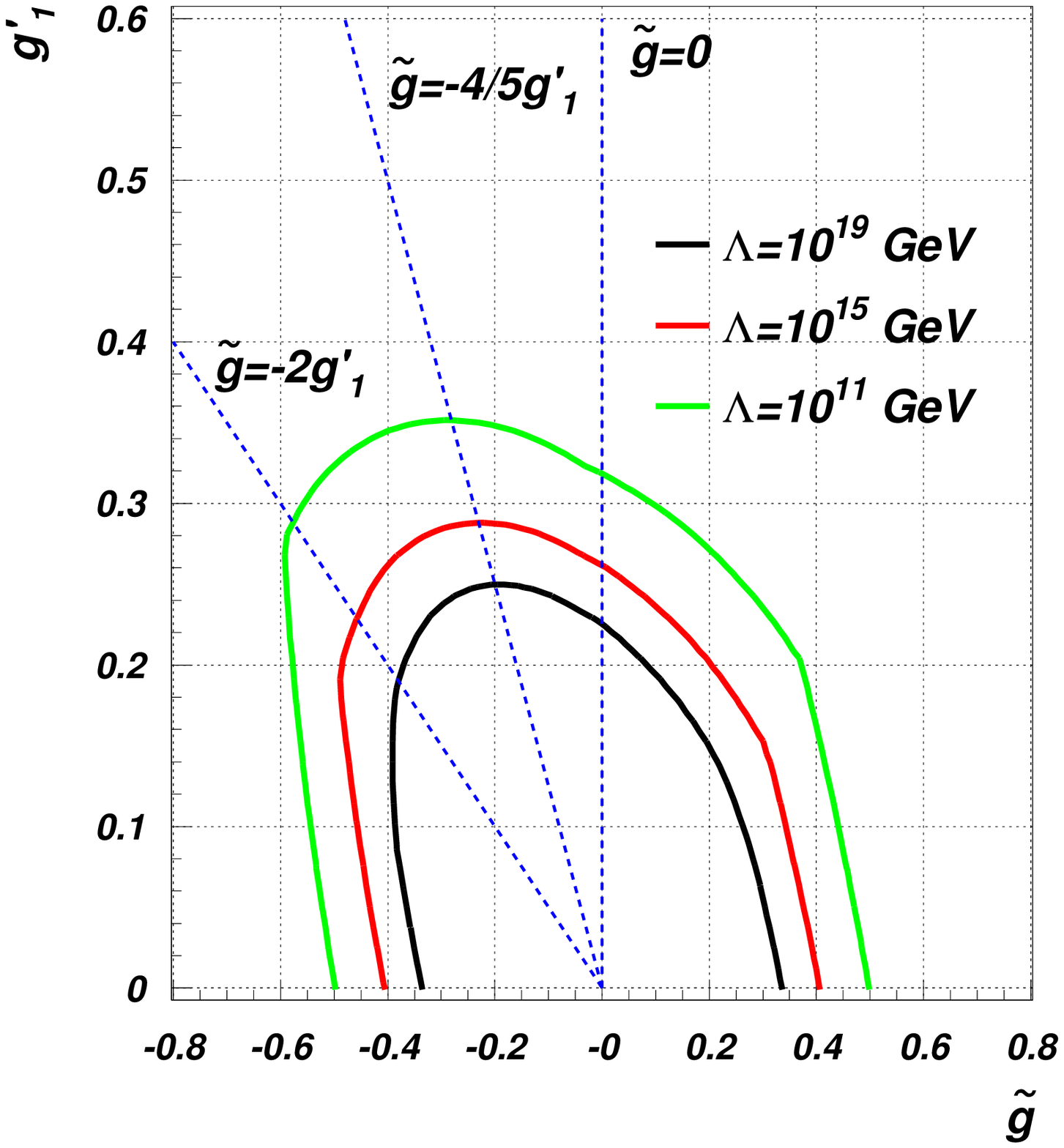}}
  \caption[]{Contour plot of both the greater eigenvalue of the spherical
  wave scattering matrix (Table~\ref{scattering-matrix}) allowed by
  unitarity (dashed lines) and the gauge couplings allowed by
  triviality (dotted lines) for several values of the cut-off energy:
  $\Lambda=10^{11}$ GeV (light green/grey lines,
  Figure~\ref{cont_11}), $\Lambda=10^{15}$ GeV (dark red/grey lines,
  Figure~\ref{cont_15}), $\Lambda=10^{19}$ GeV (black lines,
  Figure~\ref{cont_19}). Figure~\ref{cont_all} shows a summary of the
  most stringent bounds at different values of the cut-off energy,
  with focus on some peculiar parametrisation of $\widetilde{g}$
  (Table~\ref{models}).}
\label{contour}
\end{figure}
Therefore, in Figure~\ref{contour} we have overlapped the contour
plots of both the greatest 
eigenvalue of the spherical wave scattering matrix allowed by
unitarity and the gauge couplings allowed by
triviality in the $g'_1$-$\widetilde{g}$ plane for the following
values of the evolution/cut-off energy: $\Lambda=10^{11}$ GeV
(Figure~\ref{cont_11}), $\Lambda=10^{15}$ GeV (Figure~\ref{cont_15}),
$\Lambda=10^{19}$ GeV (Figure~\ref{cont_19}). It is clear that the
boundary condition imposed by the perturbative unitarity stability
(dashed lines) constrains the parameter space considerably more than
the well-known triviality bound (dotted lines). In few cases the
triviality bound is (slightly) more important than the unitarity
bound: this condition is realised at energies $\ll 10^{19}$ GeV and
$\widetilde{g}=h g'_1$ where $|h|>2$ (see
Figure~\ref{cont_11}-\ref{cont_15}). Otherwise, the unitarity condition
noticeably refines the bounds on the allowed parameter space
considerably, as it is 
clear from Figure~\ref{cont_all}, in which the most stringent bounds
are plotted 
for the aforementioned values of the cut-off energy. In the same
figure, we plotted three lines as reference for some peculiar
parametrisation of $\widetilde{g}$ already mentioned in
Section~\ref{Sec:Model}: it is clear that for each one of these models
the unitarity condition is always more important than the triviality
one.

\begin{figure}[!t]
\centering
  \subfloat[]{ 
  \label{pure}
  \includegraphics[angle=0,width=0.48\textwidth ]{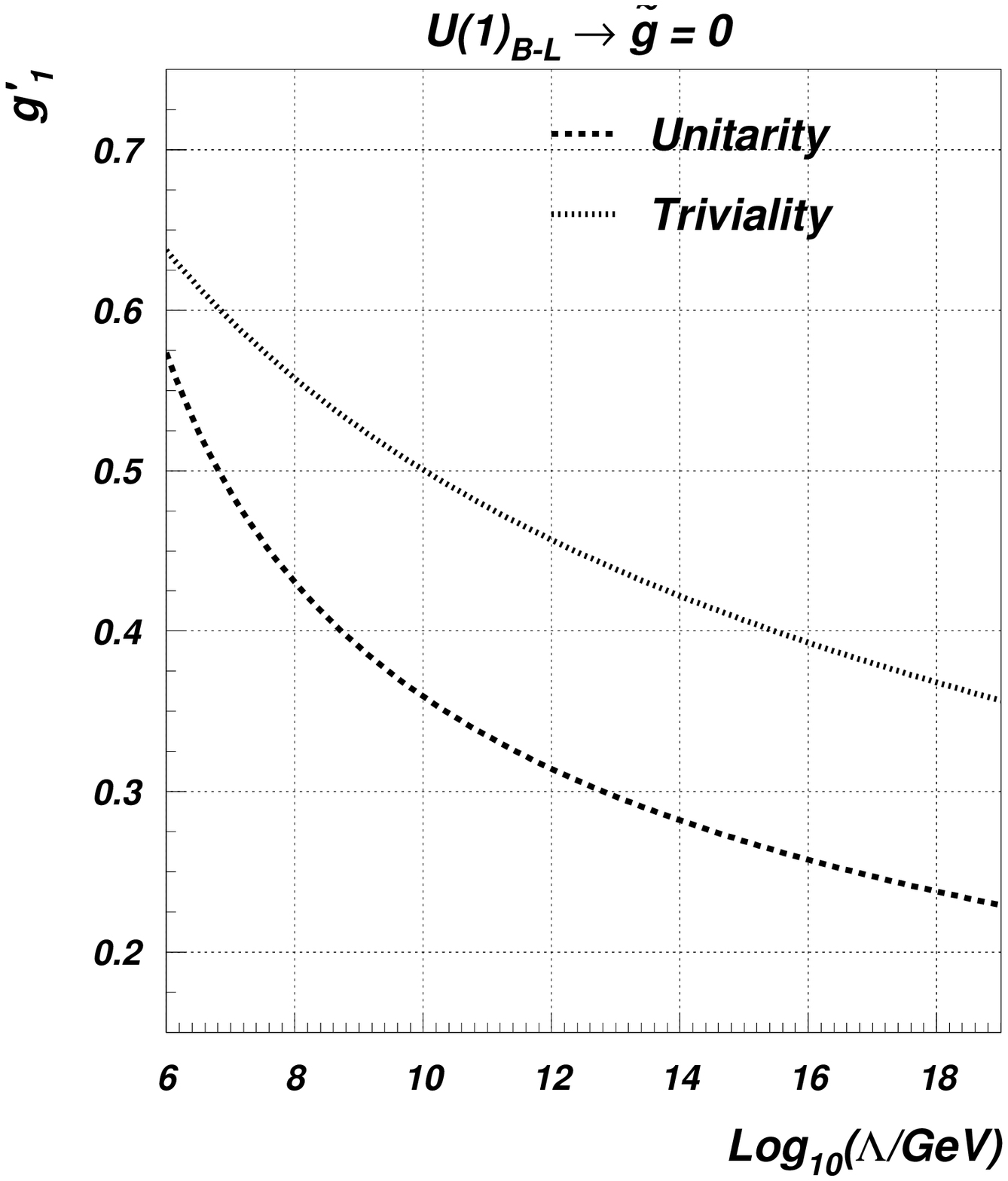}}
  \subfloat[]{ 
  \label{U1X}
  \includegraphics[angle=0,width=0.48\textwidth ]{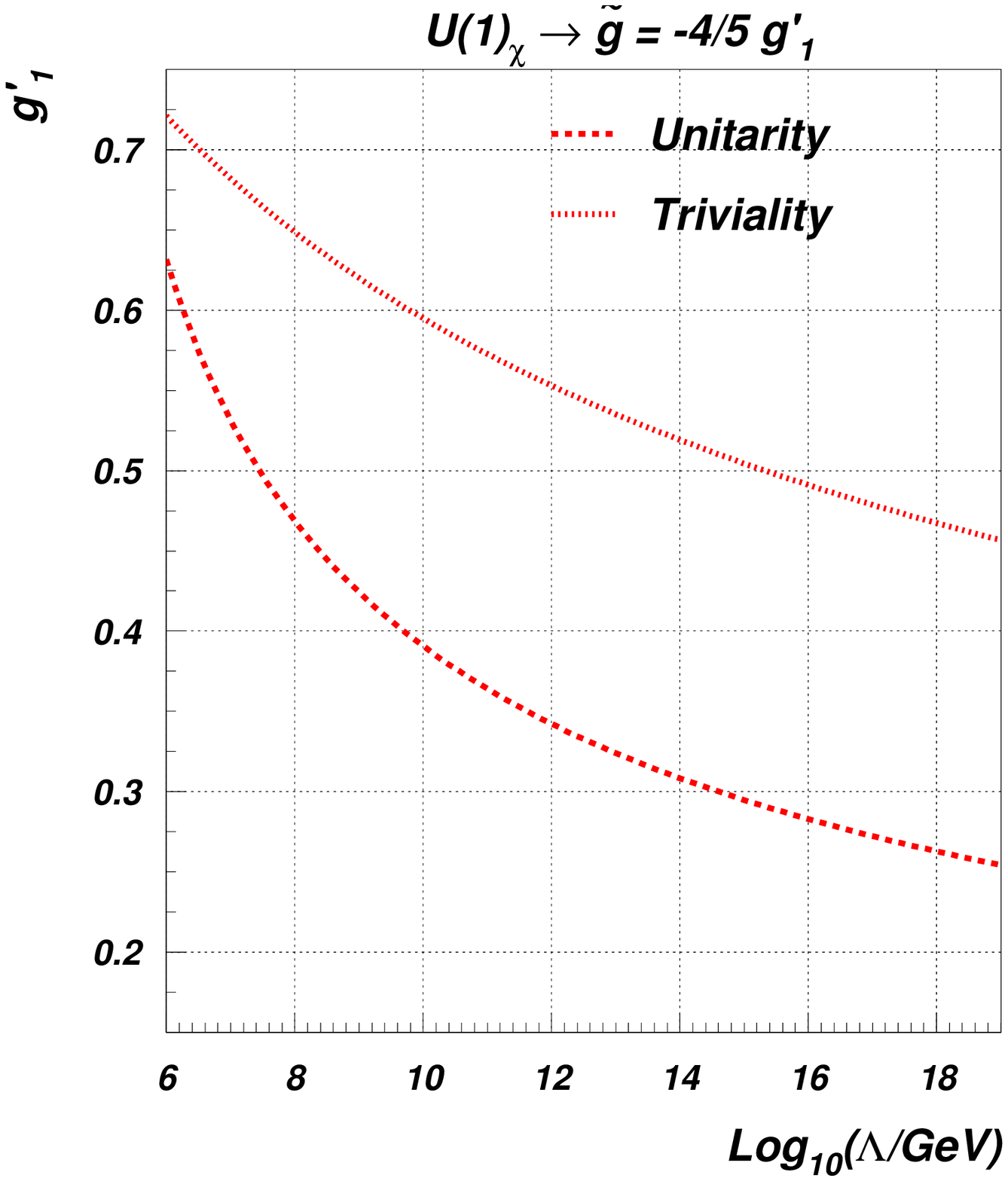}}
  \\
  \subfloat[]{
  \label{U1R}
  \includegraphics[angle=0,width=0.48\textwidth ]{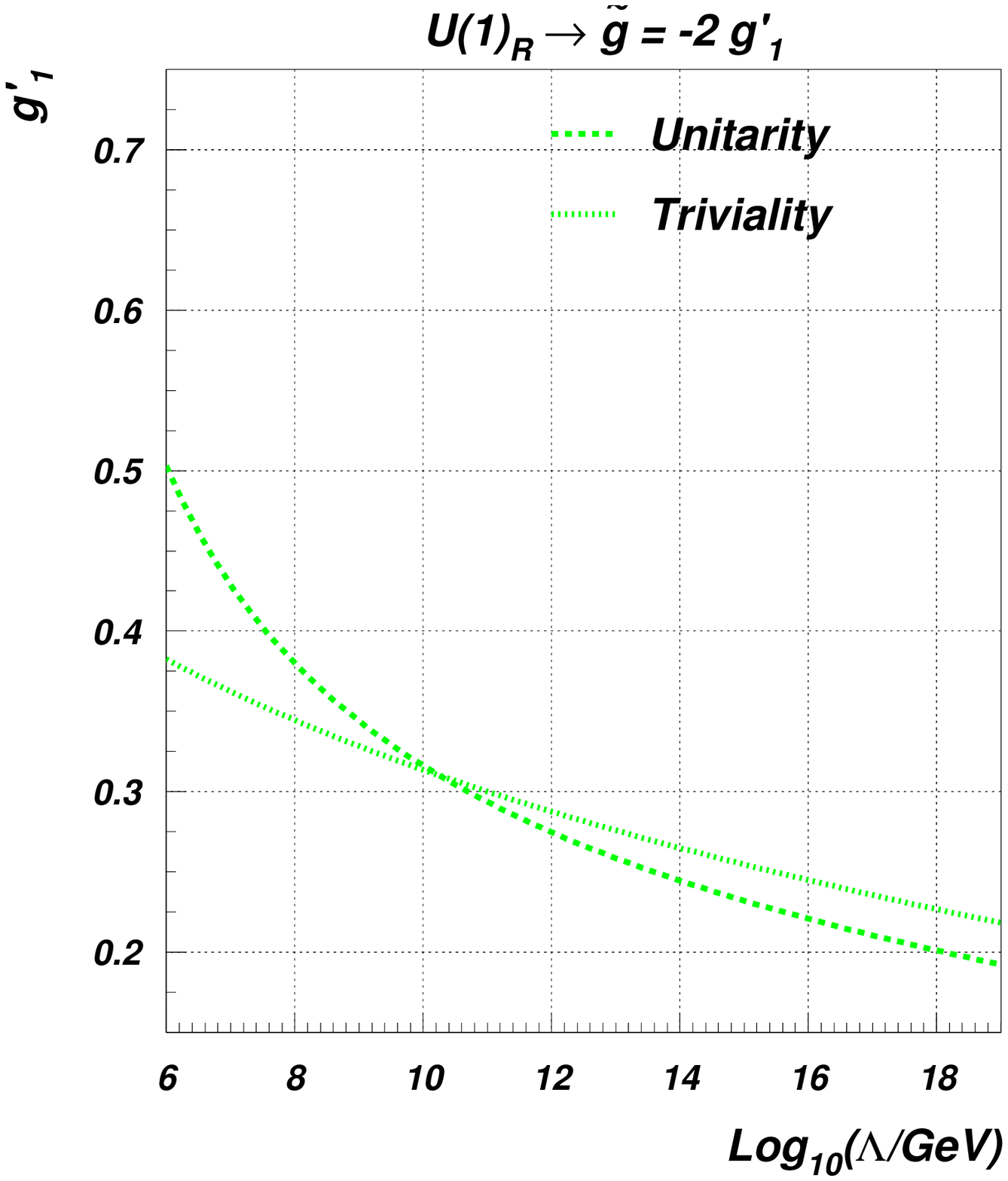}}
  \subfloat[]{
  \label{all}
  \includegraphics[angle=0,width=0.48\textwidth ]{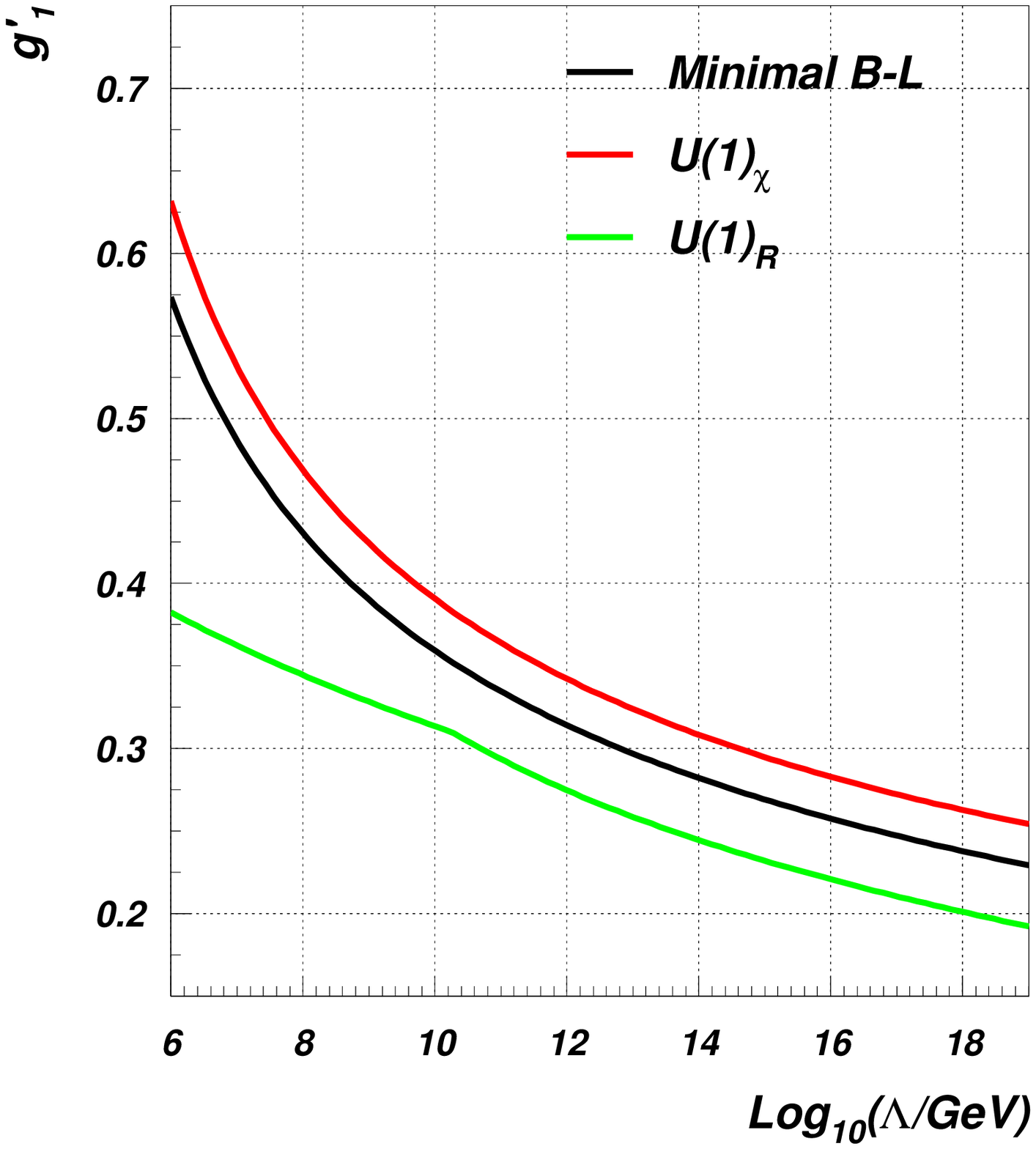}}
  \caption[]{The $g'_1$ ($\widetilde{g}$) couplings bounded either by triviality
  (dotted lines) or unitarity (dashed lines) conditions plotted
  against the evolution/cut-off energy for several peculiar choices of
  the gauge couplings parametrisation: $U(1)_{B-L}$ (black
  lines, $\widetilde{g}=0$: Figure~\ref{pure}), $U(1)_{\chi}$ (dark
  red/grey lines, $\widetilde{g}=-4/5g'_1$: Figure~\ref{U1X}), 
  $U(1)_{R}$ (light green/grey lines, $\widetilde{g}=-2g'_1$:
  Figure~\ref{U1R}). Figure~\ref{all} shows a summary of the 
  most stringent bounds at different values of the considered
  parametrisations of $\widetilde{g}$ (Table~\ref{models}).}
\label{2-D}
\end{figure}

As for these specific parametrisations, in Figure~\ref{2-D} we have
plotted the boundary value of $g'_1$ against the evolution/cut-off
scale $\Lambda$, using both the perturbative unitarity stability
condition (dashed lines) and the triviality condition (dotted
lines). For the $U(1)_{B-L}$ (Figure~\ref{pure}) and the
$U(1)_\chi$ (Figure~\ref{U1X}) extensions of the $SM$ model, the
unitarity bound is always more stringent than the triviality one. For
the $U(1)_R$ (Figure~\ref{U1R}) extension, this is only true if
$\Lambda>10^{10}$ GeV. In Figure~\ref{all} we have plotted the best
bound on $g'_1$ (and then $\widetilde{g}$) against the
evolution/cut-off energy scale $\Lambda$.

In order to summarise these results, in
Table~\ref{g1p-up_bound} we present
a comparison between the triviality and the unitarity bounds on $g'_1$
for several values of the energy scale $\Lambda$ for our choice of
Minimal $Z'$ Models.

\begin{table}[!t]
\begin{center}
\begin{tabular}{|c|c|c|c|c|c|c|c|c|c|}
\hline
\multicolumn{2}{|c|}{$Log_{10} (\Lambda/ \mbox{GeV}) $} & 7 & 9 & 11 &
13 & 15 & 17 & 19  \\
\hline
\multirow{2}{*}{$U(1)_{B-L}$}
& T & 0.594 & 0.527 & 0.477 & 0.439 & 0.407 & 0.380 & 0.357 \\ 
\ & U & 0.487 & 0.390 & 0.335 & 0.297 & 0.269 & 0.247 & 0.229 \\ 
\hline
\multirow{2}{*}{$U(1)_{\chi}$}
& T & 0.682 & 0.620 & 0.573 & 0.535 & 0.504 & 0.479 & 0.457  \\ 
\ & U & 0.531 & 0.424 & 0.364 & 0.324 & 0.295 & 0.272 & 0.254 \\ 
\hline
\multirow{2}{*}{$U(1)_{R}$}
& T & 0.362 & 0.328 & 0.300 & 0.276 & 0.254 & 0.235 & 0.218  \\ 
\ & U & 0.429 & 0.344 & 0.293 & 0.258 & 0.232 & 0.210 & 0.192 \\ 
\hline
\end{tabular}
\end{center}
\vskip -0.5cm
\caption{Triviality bounds (with $k=1$) and unitarity bounds on $g'_1$
in (non-exotic) Minimal $Z'$ Models, for several values of the
energy scale $\Lambda$.
\label{g1p-up_bound}}
\end{table}

%% file: sect_5.tex
In this paper, we have shown that, by combining perturbative unitarity and $RGE$ methods, one can significantly constrain the gauge couplings ($g'_1$ and $\widetilde{g}$) of a generic/universal (non-exotic/non-anomalous) $Z'$ gauge boson, by imposing limits on their upper values that are more stringent than standard triviality bounds.
(Also notice that, as unitarity is more constraining than triviality,
the stability of the perturbative solutions obtained through the former is already guaranteed by the latter.) 

The present work, alongside \cite{Basso:2010jt}, \cite{Basso:2010jm}
  and \cite{Basso:2010hk}, is in particular part of the long-term effort to establish the theoretical bounds on the parameter space of the $B-L$ based $U(1)$ extension of the $SM$.

%% file: acknowledgements.tex
GMP would like to thank Alexander S.~Belyaev, Stephen F.~King and
Douglas A.~Ross for helpful discussions. This work is supported in part by the NExT Institute.

%% file: appe_b.tex
As for the Goldstone bosons sector, the mass matrix and interactions
for the ghost fields are defined by the matrix $\mathcal{D}$, as in
equation~(\ref{gF}), via
\begin{eqnarray}
m^2_{ghost}=\mathcal{D}(\mathcal{D})^T.
\end{eqnarray}

Notice that the
$m^2_{ghost}$ and the $m^2_v$ of equation~(\ref{GB_mass_matrix}) have
different numbers of zero-eigenvalues, but their non-zero eigenvalues
are in a one-to-one correspondence; furthermore, the eigenvalues of
the gauge-fixing mass matrix are the same of the gauge boson mass
matrix.

Then, the ghost Lagrangian is defined, in the t'Hooft-Feynman gauge,
as 
\begin{equation}\label{L_gh}
\mathcal{L}_{ghost} = -\bar{c}^a \left[ (\partial _\mu D^\mu) ^{ab} +
\mathcal{D}^a\cdot \left(\mathcal{D}^b +\mathcal{S}^b \right)^T\right]
c^b\, ,
\end{equation}
where the matrix $\mathcal{S}$ represents the link between the
fluctuations (Goldstones) of the Higgses around their $VEVs$ and the
gauge bosons; a convenient way to write this matrix is
\begin{equation}
(\mathcal{S})^T = \left(
\begin{array}{ccccc}
\displaystyle\frac{g}{2}h & \displaystyle\frac{g}{2}\phi_3
& \displaystyle-\frac{g}{2}\phi_2 & \displaystyle-\frac{g_1}{2}\phi_2
& \displaystyle-\frac{\widetilde{g}}{2}\phi_2 \\
\displaystyle-\frac{g}{2}\phi_3 & \displaystyle \frac{g}{2}h
& \displaystyle \frac{g}{2}\phi_1 & \displaystyle \frac{g_1}{2}\phi_1
& \displaystyle \frac{\widetilde{g}}{2}\phi_1 \\
\displaystyle \frac{g}{2}\phi_2 & \displaystyle-\frac{g}{2}\phi_1
& \displaystyle \frac{g}{2}h & \displaystyle-\frac{g_1}{2}h
& \displaystyle-\frac{\widetilde{g}}{2}h \\
\displaystyle 0	& \displaystyle	0 & \displaystyle 0 & \displaystyle 0
& \displaystyle-2h'g'_1
\end{array}\right)\, .
\end{equation}

Finally, the ghost fields ($\stackrel{(-)}{c}$) read as
\begin{equation}
c = \left( \begin{array}{ccccc} 
w_1^g & w_2^g & w_3^g & B^g & (B')^g 
\end{array}\right)\, .
\end{equation}

%% file: appe_a.tex
In the following we list the set of Feynman rules that enter in the
calculation described Section~\ref{Sec:bounds}; these have been
obtained by means of implementing the information of the scalar
Lagrangian (see Section~\ref{Sec:Model} and Appendix~\ref{appe:b}) in
the LanHEP package \cite{Semenov:1996es}; all the momenta $p$'s are
considered in-coming:
\begin{eqnarray}
h_{1}-Z_{}-z_{} &:\Rightarrow &
	\frac{ -1}{ 2c_W \ s_W}\big( s_W \ s_{\theta'} \ c_W
        \ c_{\alpha} \ c_{\alpha g} \ \widetilde{g} \ p_{h}^\mu - s_W
        \ s_{\theta'} \ c_W \ c_{\alpha} \ c_{\alpha g}
        \ \widetilde{g} \ p_{z}^\mu \ \nonumber \\[2mm]
 \ &+& c_{\alpha} \ c_{\alpha g} \ c_{\theta'} \ e \ p_{z}^\mu -
 c_{\alpha} \ c_{\alpha g} \ c_{\theta'} \ e \ p_{h}^\mu -4 s_W
 \ s_{\alpha} \ s_{\alpha g} \ s_{\theta'} \ c_W \ g'_1 \ p_{h}^\mu
 \ \nonumber \\[2mm] 
 \ &+& 4 s_W \ s_{\alpha} \ s_{\alpha g} \ s_{\theta'} \ c_W \ g'_1
 \ p_{z}^\mu \big)\\[2mm] 
h_{1}-Z_{}-z'_{} &:\Rightarrow &
	\frac{ 1}{ 2c_W \ s_W}\big( s_W \ s_{\alpha g} \ s_{\theta'}
        \ c_W \ c_{\alpha} \ \widetilde{g} \ p_{h}^\mu - s_W
        \ s_{\alpha g} \ s_{\theta'} \ c_W \ c_{\alpha}
        \ \widetilde{g} \ p_{z'}^\mu \ \nonumber \\[2mm]
 \ &+& s_{\alpha g} \ c_{\alpha} \ c_{\theta'} \ e \ p_{z'}^\mu -
 s_{\alpha g} \ c_{\alpha} \ c_{\theta'} \ e \ p_{h}^\mu +4 s_W
 \ s_{\alpha} \ s_{\theta'} \ c_W \ c_{\alpha g} \ g'_1 \ p_{h}^\mu \ \nonumber \\[2mm]
 \ &-&4 s_W \ s_{\alpha} \ s_{\theta'} \ c_W \ c_{\alpha g} \ g'_1
 \ p_{z'}^\mu \big)
\end{eqnarray}
\begin{eqnarray}
h_{1}-z_{}-Z'_{} &:\Rightarrow &
        \frac{ -1}{ 2c_W \ s_W}\big( s_W \ c_W \ c_{\alpha}
        \ c_{\alpha g} \ c_{\theta'} \ \widetilde{g} \ p_{h}^\mu - s_W
        \ c_W \ c_{\alpha} \ c_{\alpha g} \ c_{\theta'}
        \ \widetilde{g} \ p_{z}^\mu \ \nonumber \\[2mm]
 \ &-& s_{\theta'} \ c_{\alpha} \ c_{\alpha g} \ e \ p_{z}^\mu +
 s_{\theta'} \ c_{\alpha} \ c_{\alpha g} \ e \ p_{h}^\mu -4 s_W
 \ s_{\alpha} \ s_{\alpha g} \ c_W \ c_{\theta'} \ g'_1 \ p_{h}^\mu
 \ \nonumber \\[2mm] 
 \ &+& 4 s_W \ s_{\alpha} \ s_{\alpha g} \ c_W \ c_{\theta'} \ g'_1
 \ p_{z}^\mu \big)\\[2mm]
h_{1}-Z'_{}-z'_{} &:\Rightarrow &
	\frac{ 1}{ 2c_W \ s_W}\big( s_W \ s_{\alpha g} \ c_W
        \ c_{\alpha} \ c_{\theta'} \ \widetilde{g} \ p_{h}^\mu - s_W
        \ s_{\alpha g} \ c_W \ c_{\alpha} \ c_{\theta'}
        \ \widetilde{g} \ p_{z'}^\mu \nonumber \\[2mm]
 \ &-& s_{\alpha g} \ s_{\theta'} \ c_{\alpha} \ e \ p_{z'}^\mu +
 s_{\alpha g} \ s_{\theta'} \ c_{\alpha} \ e \ p_{h}^\mu +4 s_W
 \ s_{\alpha} \ c_W \ c_{\alpha g} \ c_{\theta'} \ g'_1 \ p_{h}^\mu
 \ \nonumber \\[2mm] 
 \ &-&4 s_W \ s_{\alpha} \ c_W \ c_{\alpha g} \ c_{\theta'} \ g'_1
 \ p_{z'}^\mu \big)
\end{eqnarray}
\begin{eqnarray}
h_{2}-Z_{}-z_{} &:\Rightarrow &
	\frac{ -1}{ 2c_W \ s_W}\big( s_W \ s_{\alpha} \ s_{\theta'}
        \ c_W \ c_{\alpha g} \ \widetilde{g} \ p_{h}^\mu - s_W
        \ s_{\alpha} \ s_{\theta'} \ c_W \ c_{\alpha g}
        \ \widetilde{g} \ p_{z}^\mu \ \nonumber \\[2mm]
 \ &+& s_{\alpha} \ c_{\alpha g} \ c_{\theta'} \ e \ p_{z}^\mu -
 s_{\alpha} \ c_{\alpha g} \ c_{\theta'} \ e \ p_{h}^\mu +4 s_W
 \ s_{\alpha g} \ s_{\theta'} \ c_W \ c_{\alpha} \ g'_1 \ p_{h}^\mu
 \ \nonumber \\[2mm] 
 \ &-&4 s_W \ s_{\alpha g} \ s_{\theta'} \ c_W \ c_{\alpha} \ g'_1
 \ p_{z}^\mu \big)\\[2mm]
h_{2}-Z_{}-z'_{} &:\Rightarrow &
	\frac{ 1}{ 2c_W \ s_W}\big( s_W \ s_{\alpha} \ s_{\alpha g}
        \ s_{\theta'} \ c_W \ \widetilde{g} \ p_{h}^\mu - s_W
        \ s_{\alpha} \ s_{\alpha g} \ s_{\theta'} \ c_W
        \ \widetilde{g} \ p_{z'}^\mu \nonumber \\[2mm]
 \ &+& s_{\alpha} \ s_{\alpha g} \ c_{\theta'} \ e \ p_{z'}^\mu -
 s_{\alpha} \ s_{\alpha g} \ c_{\theta'} \ e \ p_{h}^\mu -4 s_W
 \ s_{\theta'} \ c_W \ c_{\alpha} \ c_{\alpha g} \ g'_1 \ p_{h}^\mu
 \ \nonumber \\[2mm] 
 \ &+&4 s_W \ s_{\theta'} \ c_W \ c_{\alpha} \ c_{\alpha g} \ g'_1
 \ p_{z'}^\mu \big)
\end{eqnarray}
\begin{eqnarray} 
h_{2}-z_{}-Z'_{} &:\Rightarrow &
	\frac{ -1}{ 2c_W \ s_W}\big( s_W \ s_{\alpha} \ c_W
        \ c_{\alpha g} \ c_{\theta'} \ \widetilde{g} \ p_{h}^\mu - s_W
        \ s_{\alpha} \ c_W \ c_{\alpha g} \ c_{\theta'}
        \ \widetilde{g} \ p_{z}^\mu \ \nonumber \\[2mm]
 \ &-& s_{\alpha} \ s_{\theta'} \ c_{\alpha g} \ e \ p_{z}^\mu +
 s_{\alpha} \ s_{\theta'} \ c_{\alpha g} \ e \ p_{h}^\mu +4 s_W
 \ s_{\alpha g} \ c_W \ c_{\alpha} \ c_{\theta'} \ g'_1 \ p_{h}^\mu
 \ \nonumber \\[2mm] 
 \ &-&4 s_W \ s_{\alpha g} \ c_W \ c_{\alpha} \ c_{\theta'} \ g'_1
 \ p_{z}^\mu \big)\\[2mm] 
h_{2}-Z'_{}-z'_{} &:\Rightarrow &
	\frac{ 1}{ 2c_W \ s_W}\big( s_W \ s_{\alpha} \ s_{\alpha g}
        \ c_W \ c_{\theta'} \ \widetilde{g} \ p_{h}^\mu - s_W
        \ s_{\alpha} \ s_{\alpha g} \ c_W \ c_{\theta'}
        \ \widetilde{g} \ p_{z'}^\mu \ \nonumber \\[2mm]
 \ &-& s_{\alpha} \ s_{\alpha g} \ s_{\theta'} \ e \ p_{z'}^\mu +
 s_{\alpha} \ s_{\alpha g} \ s_{\theta'} \ e \ p_{h}^\mu -4 s_W \ c_W
 \ c_{\alpha} \ c_{\alpha g} \ c_{\theta'} \ g'_1 \ p_{h}^\mu
 \ \nonumber \\[2mm] 
 \ &+&4 s_W \ c_W \ c_{\alpha} \ c_{\alpha g} \ c_{\theta'} \ g'_1
 \ p_{z'}^\mu \big)
\end{eqnarray}
\begin{eqnarray}
w^+_{}-w^-_{}-Z_{} &:\Rightarrow &
	\frac{i}{ 2c_W \ s_W}\big( (1-2 s_W^2) \ c_{\theta'} \ e \
        p_{w^-}^\mu + s_W \ s_{\theta'} \ c_W \ \widetilde{g} \
        p_{w^-}^\mu \ \nonumber \\[2mm] 
 \ &-& (1-2 s_W {}^2) \ c_{\theta'} \ e \ p_{w^+}^\mu - s_W
 \ s_{\theta'} \ c_W \ \widetilde{g} \ p_{w^+}^\mu \big)\\[2mm] 
w^+_{}-w^-_{}-Z'_{} &:\Rightarrow &
	\frac{-i}{ 2c_W \ s_W}\big( (1-2 s_W^2) \ s_{\theta'} \ e \ 
        p_{w^-}^\mu - s_W \ c_W \ c_{\theta'} \ \widetilde{g} \
        p_{w^-}^\mu \ \nonumber \\[2mm] 
 \ &-& (1-2 s_W^2) \ s_{\theta'} \ e \ p_{w^+}^\mu + s_W \ c_W
 \ c_{\theta'} 
 \ \widetilde{g} \ p_{w^+}^\mu \big)
\end{eqnarray}

In the previous formulae, we have used the following notation:
\begin{eqnarray}
c_W(s_W) &\rightarrow& \cos{\theta_W}(\sin{\theta_W}), \nonumber \\
c_{\alpha}(s_{\alpha}) &\rightarrow& \cos{\alpha}(\sin{\alpha}), \nonumber \\
c_{\alpha g}(s_{\alpha
g}) &\rightarrow& \cos{\alpha_g}(\sin{\alpha_g}),\label{notation} \\ 
c_{\theta'}(s_{\theta'}) &\rightarrow&
\cos{\theta'}(\sin{\theta'}),\nonumber \\
e&\rightarrow&\frac{gg_1}{\sqrt{g^2+g_1^2}}. \nonumber
\end{eqnarray}

%% file: appe_c.tex
The non-vanishing coefficients related to the structure of
equation~(\ref{structure}), in the high energy limit, for each entry
of the scattering matrix are the following: 
\begin{eqnarray}
&\ & f^{z}_{zz,h_1h_1} =\nonumber \\
&=& \frac{1}{4} \left(-c_{\alpha}^2 c_{\alpha g}^2
c_{\theta'}^2 \left(g^2+g_1^2\right)+(c_{\alpha} c_{\alpha
g} \widetilde{g}-4 g'_1 s_{\alpha} s_{\alpha g})^2
s_{\theta'}^2\right), \\[2mm]
&\ & f^{z'}_{zz,h_1h_1}=\nonumber \\
&=& \frac{1}{4} \left(16 c_{\theta'}^2 (g'_1)^2
s_{\alpha}^2 s_{\alpha g}^2-8 c_{\alpha} c_{\alpha g} c_{\theta'} g'_1
s_{\alpha} s_{\alpha
g} \left(c_{\theta'} \widetilde{g}+\sqrt{g^2+g_1^2}
s_{\theta'}\right)\right. \nonumber \\
&+& \left. c_{\alpha}^2 c_{\alpha
g}^2 \left(c_{\theta'}^2 \widetilde{g}^2+2
c_{\theta'} \sqrt{g^2+g_1^2} \widetilde{g}
s_{\theta'}+\left(g^2+g_1^2\right) s_{\theta'}^2\right)\right), \\[2mm]
&\ & f^{z}_{zz,h_1h_2}=\nonumber \\
&=& \frac{1}{4} \left(4 c_{\alpha g} g'_1
s_{\alpha}^2 s_{\alpha g}
s_{\theta'} \left(c_{\theta'} \sqrt{g^2+g_1^2}-\widetilde{g}
s_{\theta'}\right)\right. \nonumber \\
&+& \left. 4 c_{\alpha}^2 c_{\alpha g} g'_1 s_{\alpha g}
s_{\theta'} \left(-c_{\theta'} \sqrt{g^2+g_1^2}+\widetilde{g}
s_{\theta'}\right)\right. \nonumber \\
&+& \left. c_{\alpha} s_{\alpha} \left(-16 (g'_1)^2 s_{\alpha
g}^2 s_{\theta'}^2+c_{\alpha
g}^2 \left(c_{\theta'}^2 \left(g^2+g_1^2\right)-2
c_{\theta'} \sqrt{g^2+g_1^2} \widetilde{g} s_{\theta'}+\widetilde{g}^2
s_{\theta'}^2\right)\right)\right), \\ [2mm]
&\ & f^{z'}_{zz,h_1h_2}=\nonumber \\
&=& \frac{1}{4} \left(4 c_{\alpha}^2 c_{\alpha g}
c_{\theta'} g'_1 s_{\alpha
g} \left(c_{\theta'} \widetilde{g}+\sqrt{g^2+g_1^2}
s_{\theta'}\right)\right. \nonumber \\
&-& \left. 4 c_{\alpha g} c_{\theta'} g'_1 s_{\alpha}^2
s_{\alpha g} \left(c_{\theta'} \widetilde{g}+\sqrt{g^2+g_1^2}
s_{\theta'}\right)\right. \nonumber \\
&+&\left. c_{\alpha} s_{\alpha} \left(-16 c_{\theta'}^2
(g'_1)^2 s_{\alpha g}^2+c_{\alpha
g}^2 \left(c_{\theta'}^2 \widetilde{g}^2+2
c_{\theta'} \sqrt{g^2+g_1^2} \widetilde{g}
s_{\theta'}+\left(g^2+g_1^2\right)
s_{\theta'}^2\right)\right)\right), \\  [2mm]
&\ & f^{z}_{zz,h_2h_2}=\nonumber \\
&=& \frac{1}{4} \left(c_{\alpha g}^2
c_{\theta'}^2 \left(g^2+g_1^2\right) s_{\alpha}^2-2 c_{\alpha g}^2
c_{\theta'} \sqrt{g^2+g_1^2} \widetilde{g} s_{\alpha}^2
s_{\theta'}\right. \nonumber \\
&-&\left. 8 c_{\alpha} c_{\alpha g} c_{\theta'} \sqrt{g^2+g_1^2} g'_1
s_{\alpha} s_{\alpha g} s_{\theta'}+(c_{\alpha g} \widetilde{g} s_{\alpha}+4
c_{\alpha} g'_1 s_{\alpha g})^2 s_{\theta'}^2\right)\\  [2mm]
&\ & f^{z'}_{zz,h_2h_2}=\nonumber \\
&=& \frac{1}{4} \left(c_{\theta'}^2 (c_{\alpha
g} \widetilde{g} s_{\alpha}+4 c_{\alpha} g'_1 s_{\alpha
g})^2-c_{\alpha g}^2 \left(g^2+g_1^2\right) s_{\alpha}^2
s_{\theta'}^2\right),
\end{eqnarray}
\begin{eqnarray}
&\ & f^{z}_{zz',h_1h_1} =\nonumber \\
&=& \frac{1}{4} \left(16 c_{\alpha g} (g'_1)^2
s_{\alpha}^2 s_{\alpha g} s_{\theta'}^2+4 c_{\alpha} g'_1
s_{\alpha} \left(c_{\alpha g}^2+s_{\alpha g}^2\right)
s_{\theta'} \left(c_{\theta'} \sqrt{g^2+g_1^2}-\widetilde{g}
s_{\theta'}\right)\right. \nonumber \\
&+&\left. c_{\alpha}^2 c_{\alpha g} s_{\alpha
g} \left(c_{\theta'}^2 \left(g^2+g_1^2\right)-2
c_{\theta'} \sqrt{g^2+g_1^2} \widetilde{g} s_{\theta'}+\widetilde{g}^2
s_{\theta'}^2\right)\right), \\ [2mm]
&\ & f^{z'}_{zz',h_1h_1} =\nonumber \\
&=& \frac{1}{4} \left(16 c_{\alpha g}
c_{\theta'}^2 (g'_1)^2 s_{\alpha}^2 s_{\alpha g}-4 c_{\alpha}
c_{\theta'} g'_1 s_{\alpha} \left(c_{\alpha g}^2+s_{\alpha
g}^2\right) \left(c_{\theta'} \widetilde{g}+\sqrt{g^2+g_1^2}
s_{\theta'}\right)\right. \nonumber \\
&+&\left. c_{\alpha}^2 c_{\alpha g} s_{\alpha
g} \left(c_{\theta'}^2 \widetilde{g}^2+2
c_{\theta'} \sqrt{g^2+g_1^2} \widetilde{g}
s_{\theta'}+\left(g^2+g_1^2\right) s_{\theta'}^2\right)\right), \\ [2mm]
&\ & f^{z}_{zz',h_1h_2} =\nonumber \\ 
&=& \frac{1}{4} \left(4 g'_1 s_{\alpha}^2 s_{\alpha
g}^2 s_{\theta'} \left(c_{\theta'} \sqrt{g^2+g_1^2}-\widetilde{g}
s_{\theta'}\right)+4 c_{\alpha}^2 c_{\alpha g}^2 g'_1
s_{\theta'} \left(-c_{\theta'} \sqrt{g^2+g_1^2}+\widetilde{g}
s_{\theta'}\right)\right. \nonumber \\
&+&\left. c_{\alpha} c_{\alpha g} s_{\alpha} s_{\alpha
g} \left(c_{\theta'}^2 \left(g^2+g_1^2\right)-2
c_{\theta'} \sqrt{g^2+g_1^2} \widetilde{g} s_{\theta'}+\left(-16
(g'_1)^2+\widetilde{g}^2\right) s_{\theta'}^2\right)\right), \\ [2mm]
&\ & f^{z'}_{zz',h_1h_2}  =\nonumber \\
&=& \frac{1}{4} \left(4 c_{\alpha}^2 c_{\alpha
g}^2 c_{\theta'} g'_1 \left(c_{\theta'} \widetilde{g}+\sqrt{g^2+g_1^2}
s_{\theta'}\right)-4 c_{\theta'} g'_1 s_{\alpha}^2 s_{\alpha
g}^2 \left(c_{\theta'} \widetilde{g}+\sqrt{g^2+g_1^2}
s_{\theta'}\right)\right. \nonumber \\
&+&\left. c_{\alpha} c_{\alpha g} s_{\alpha} s_{\alpha
g} \left(c_{\theta'}^2 \left(-16 (g'_1)^2+\widetilde{g}^2\right)+2
c_{\theta'} \sqrt{g^2+g_1^2} \widetilde{g}
s_{\theta'}+\left(g^2+g_1^2\right) s_{\theta'}^2\right)\right), \\ [2mm]
&\ & f^{z}_{zz',h_2h_2}  =\nonumber \\
&=& \frac{1}{4} \left(4 c_{\alpha} g'_1 s_{\alpha}
s_{\alpha g}^2
s_{\theta'} \left(c_{\theta'} \sqrt{g^2+g_1^2}-\widetilde{g}
s_{\theta'}\right)-4 c_{\alpha} c_{\alpha g}^2 g'_1 s_{\alpha}
s_{\theta'} \left(c_{\theta'} \sqrt{g^2+g_1^2}-\widetilde{g}
s_{\theta'}\right)\right. \nonumber \\
&+&\left. c_{\alpha g} s_{\alpha
g} \left(c_{\theta'}^2 \left(g^2+g_1^2\right) s_{\alpha}^2-2
c_{\theta'} \sqrt{g^2+g_1^2} \widetilde{g} s_{\alpha}^2
s_{\theta'}+\left(-16 c_{\alpha}^2 (g'_1)^2+\widetilde{g}^2
s_{\alpha}^2\right) s_{\theta'}^2\right)\right), \\ [2mm]
&\ & f^{z'}_{zz',h_2h_2} =\nonumber \\ 
&=&\frac{1}{4} \left(-16 c_{\alpha}^2 c_{\alpha g} 
c_{\theta'}^2 (g'_1)^2 s_{\alpha g}+4 c_{\alpha} c_{\theta'} g'_1
s_{\alpha} (c_{\alpha g}^2-s_{\alpha
g}^2) \left(c_{\theta'} \widetilde{g}+\sqrt{g^2+g_1^2}
s_{\theta'}\right)\right. \nonumber \\
&+&\left. c_{\alpha g} s_{\alpha}^2 s_{\alpha
g} \left(c_{\theta'}^2 \widetilde{g}^2+2
c_{\theta'} \sqrt{g^2+g_1^2} \widetilde{g}
s_{\theta'}+\left(g^2+g_1^2\right) s_{\theta'}^2\right)\right),
\end{eqnarray}
\begin{eqnarray}
&\ & f^{z}_{z'z',h_1h_1} =\nonumber \\
&=& \frac{1}{4} \left(c_{\alpha}^2
c_{\theta'}^2 \left(g^2+g_1^2\right) s_{\alpha g}^2+8 c_{\alpha}
c_{\alpha g} c_{\theta'} \sqrt{g^2+g_1^2} g'_1 s_{\alpha} s_{\alpha g}
s_{\theta'}\right. \nonumber \\
&-&\left. 2 c_{\alpha}^2 c_{\theta'} \sqrt{g^2+g_1^2} \widetilde{g}
s_{\alpha g}^2 s_{\theta'}+(-4 c_{\alpha g} g'_1
s_{\alpha}+c_{\alpha} \widetilde{g} s_{\alpha g})^2
s_{\theta'}^2\right), \\ [2mm]
&\ & f^{z'}_{z'z',h_1h_1} =\nonumber \\
&=& \frac{1}{4} \left(16 c_{\alpha g}^2 c_{\theta'}^2 (g'_1)^2
s_{\alpha}^2-8 c_{\alpha} c_{\alpha g} c_{\theta'} g'_1 s_{\alpha}
s_{\alpha g} \left(c_{\theta'} \widetilde{g}+\sqrt{g^2+g_1^2}
s_{\theta'}\right)\right. \nonumber \\
&+&\left. c_{\alpha}^2 s_{\alpha
g}^2 \left(c_{\theta'}^2 \widetilde{g}^2+2
c_{\theta'} \sqrt{g^2+g_1^2} \widetilde{g}
s_{\theta'}+\left(g^2+g_1^2\right) s_{\theta'}^2\right)\right), \\ [2mm]
&=& f^{z}_{z'z',h_1h_2} =\nonumber \\
&=& \frac{1}{4} \left(4 c_{\alpha g} g'_1
s_{\alpha}^2 s_{\alpha g}
s_{\theta'} \left(c_{\theta'} \sqrt{g^2+g_1^2}-\widetilde{g}
s_{\theta'}\right)\right. \nonumber \\
&+&\left. 4 c_{\alpha}^2 c_{\alpha g} g'_1 s_{\alpha g}
s_{\theta'} \left(-c_{\theta'} \sqrt{g^2+g_1^2}+\widetilde{g}
s_{\theta'}\right)\right. \nonumber \\
&+&\left. c_{\alpha}
s_{\alpha} \left(c_{\theta'}^2 \left(g^2+g_1^2\right) s_{\alpha g}^2-2
c_{\theta'} \sqrt{g^2+g_1^2} \widetilde{g} s_{\alpha g}^2
s_{\theta'}-\left(16 c_{\alpha g}^2 (g'_1)^2-\widetilde{g}^2
s_{\alpha g}^2\right) s_{\theta'}^2\right)\right), \\ [2mm]
&\ &f^{z'}_{z'z',h_1h_2} =\nonumber \\
&=& \frac{1}{4} \left(4 c_{\alpha}^2 c_{\alpha g}
c_{\theta'} g'_1 s_{\alpha
g} \left(c_{\theta'} \widetilde{g}+\sqrt{g^2+g_1^2}
s_{\theta'}\right) \right. \nonumber \\
&-& 4 c_{\alpha g} c_{\theta'} g'_1 s_{\alpha}^2
s_{\alpha g} \left(c_{\theta'} \widetilde{g}+\sqrt{g^2+g_1^2}
s_{\theta'}\right) \nonumber \\
&+&\left. c_{\alpha} s_{\alpha} \left(-16 c_{\alpha g}^2
c_{\theta'}^2 (g'_1)^2+s_{\alpha
g}^2 \left(c_{\theta'}^2 \widetilde{g}^2+2
c_{\theta'} \sqrt{g^2+g_1^2} \widetilde{g}
s_{\theta'}+\left(g^2+g_1^2\right)
s_{\theta'}^2\right)\right)\right), \\ [2mm]
&\ &f^{z}_{z'z',h_2h_2}=\nonumber \\
&=& \frac{1}{4} \left(c_{\theta'}^2 \left(g^2+g_1^2\right)
s_{\alpha}^2 s_{\alpha g}^2-2 c_{\theta'} \sqrt{g^2+g_1^2} s_{\alpha}
s_{\alpha g} (4 c_{\alpha} c_{\alpha g} g'_1+\widetilde{g} s_{\alpha}
s_{\alpha g}) s_{\theta'}\right. \nonumber \\
&+&\left. (4 c_{\alpha} c_{\alpha g}
g'_1+\widetilde{g} s_{\alpha} s_{\alpha g})^2 s_{\theta'}^2\right), \\ 
&\ &f^{z'}_{z'z',h_2h_2} =\nonumber \\
&=& \frac{1}{4} \left(16 c_{\alpha}^2 c_{\alpha
g}^2 c_{\theta'}^2 (g'_1)^2+8 c_{\alpha} c_{\alpha g} c_{\theta'} g'_1
s_{\alpha} s_{\alpha
g} \left(c_{\theta'} \widetilde{g}+\sqrt{g^2+g_1^2}
s_{\theta'}\right)\right. \nonumber \\
&+&\left. s_{\alpha}^2 s_{\alpha
g}^2 \left(c_{\theta'}^2 \widetilde{g}^2+2
c_{\theta'} \sqrt{g^2+g_1^2} \widetilde{g}
s_{\theta'}+\left(g^2+g_1^2\right) s_{\theta'}^2\right)\right).
\end{eqnarray}

The non-vanishing coefficients related to the structure of
equation~(\ref{w's}), in the high energy limit, are the following:
\begin{eqnarray}
&\ &f^{z}_{w^+w^-}=\nonumber \\
&=& \frac{c_{\theta'}^2 \left(g^2-g_1^2\right)^2 \sqrt{g^2+g_1^2}+2
c_{\theta'} \left(g^4-g_1^4\right) \widetilde{g}
s_{\theta'}+\left(g^2+g_1^2\right)^{3/2} \widetilde{g}^2
s_{\theta'}^2}{4 \left(g^2+g_1^2\right)^{3/2}}, \\
&\ &f^{z'}_{w^+w^-}=\nonumber \\
&=& \frac{ c_{\theta'}^2 \left(g^2 +
g_1^2\right)^{3/2} \widetilde{g}^2-2
c_{\theta'} \left(g^4-g_1^4\right) \widetilde{g}
s_{\theta'}+\left(g^2-g_1^2\right)^2 \sqrt{g^2+g_1^2}
s_{\theta'}^2 }{4 \left(g^2 +
g_1^2\right)^{3/2}} .
\end{eqnarray}
 
In the previous equations, we have used the notation of
equations~(\ref{notation}).